\documentclass[preprint]{aastex}
\usepackage{graphicx,times}
\usepackage{natbib}
\usepackage{rotating}

\def\kms{$\rm{km~s}^{-1}$}
\small

\shorttitle{Muticolor Photometry of the Regular Galaxy Cluster A2589}
\shortauthors{Shunfang Liu et al.}

\begin{document}

\title{Multicolor Photometry Study of the Galaxy Cluster A2589: Dynamics, Luminosity Function and Star Formation History }

\author{Shun-Fang Liu\altaffilmark{1,2,3},
  Qi-Rong Yuan\altaffilmark{4},
  Yan-Bin Yang\altaffilmark{1,3},
  Jun Ma\altaffilmark{1,3},
  Zhao-Ji Jiang\altaffilmark{1,3},
  Jiang-Hua Wu\altaffilmark{1,3},
  Zhen-Yu Wu\altaffilmark{1,3},
  Jian-Sheng Chen\altaffilmark{1,3},
  Xu Zhou\altaffilmark{1,3}}
\altaffiltext{1} {National Astronomical Observatories, Chinese Academy of
Sciences, Beijing 100012, China}

\altaffiltext{2}{Graduate University, Chinese Academy of Sciences,
  Beijing, 100039, China}

\altaffiltext{3}{Key Laboratory of Optical Astronomy, National Astronomical Observatories, Chinese Academy of Sciences}

\altaffiltext{4}{Department of Physics, Nanjing Normal University, Nanjing , China}

 \email{liusf@bac.pku.edu.cn; yuanqirong@njnu.edu.cn; zhouxu@bao.ac.cn}
\vspace{0pt}
\begin{abstract}

Smooth X-ray morphology and non-detection of radio source at
center indicate that A2589 is a typical case of well-relaxed
regular galaxy cluster. In this paper we present a multicolor
photometry for A2589 ($z=0.0414$) with 15 intermediate bands in
the Beijing-Arizona-Taiwan-Connecticut (BATC) system which covers
an optical wavelength range from 3000~\AA\ to 10000~\AA. The
spectral energy distributions (SEDs) for more than 5000 sources
are achieved down to {\it V} $\sim$ 20 mag in about 1 deg$^{2}$
field. A2589 has been also covered by the Sloan Digital Sky Survey
(SDSS) in photometric mode only. A cross-identification of the
BATC-detected galaxies with the SDSS photometric catalog achieves
1199 galaxies brighter than $i=19.5$ mag, among which 68 member
galaxies with known spectroscopic redshifts are found. After
combining the SDSS five-band photometric data and the BATC SEDs,
the technique of photometric redshift is applied to these galaxies
for selecting faint member galaxies. The color-magnitude relation
is taken as a further restriction of early-type cluster galaxies.
As a result, 106 galaxies are newly selected as member galaxies.
Spatial distribution of member galaxies shows a north-south
elongation which agrees with the X-ray brightness profile and the
orientation of central cD galaxy, NGC~7647. No substructures are
detected on the basis of positions and radial velocities of
cluster galaxies, indicating that A2589 is a well-relaxed system.
The luminosity function of A2589 exhibits a peak at $M_{R} \sim
-20$ mag and a dip at $M_{R} \sim -19 $ mag. The low-density outer
regions are the preferred habitat of faint galaxies. With the
evolutionary population synthesis model, PEGASE, the environmental
effect on the star formation properties for 68 spectroscopically
confirmed member galaxies is studied. The outlier faint galaxies
tend to have longer time scales of star formation, shorter mean
stellar ages, and lower metallicities of interstellar medium,
which can be interpreted in the context of hierarchical
cosmological scenario.
\end{abstract}


\section{Introduction}
\label{Introduction.sec}

As the largest gravitationally bound systems in the universe,
galaxy clusters play a central role in cosmological studies, such
as the baryon content of the universe, formation of large-scale
structure, and the density parameters in cosmological models
\citep{bahcall96}. For instance, the distribution of cluster
velocity dispersions and mass-to-light ratios might provide limits
on the amplitude of primordial density fluctuations in
hierarchical clustering scenarios \citep{evrard89, frenk90}. In
the cold dark matter models, galaxy clusters are complex
multicomponent systems where galaxies, hot gas, and dark matter
evolve in a tightly coupled way \citep{frenk96}, and they are
driven by accreting these components along filaments
\citep{bekki01, roettiger96,navarro94, cen94}. Therefore, cluster
of galaxies has been regarded as a powerful laboratory for
studying the evolution of the galaxies in a dense environment
where the physical properties of galaxies might have been
influenced by many different mechanisms, including strong
galaxy-galaxy interaction \citep{mihos04}, harassment
\citep{moore98}, ram pressure stripping \citep{quilis00}, and
strangulation \citep[also known as starvation, or
suffocation]{bower04}.

According to the hierarchical clustering scenario, galaxy clusters
continue to be assembled, and relaxation is not yet complete in most
clusters. It is widely appreciated that a regular cluster with no
detectable substructure both in spatial distribution and in velocity
space is at the most evolved stage. However, significant
substructures have been revealed in a large number of rich galaxy
clusters with the X-ray imaging and optical spectroscopic surveys
\citep{rhee91, forman82,beers91,sarazin92,henry93,burns94}. Numerical
simulation of the evolution of galaxy clusters indicates that at
least 50\% of apparently relaxed clusters contain significant
substructures \citep{salva93}. Very few clusters with single central
dominant cD galaxy are found to be spherically symmetric in
distributions of galaxies and hot gas. Even for the Coma cluster that
was once regarded as an exemplar of regular and well-relaxed cluster,
the projected distributions of galaxies and the X-ray-emitting gas
also show convincing evidence of significant substructures on large
and small scales \citep{colless96, white93, neumann03}. Recently, a
deep spectroscopic survey for the faint galaxies in the central
region of Coma cluster has confirmed previously identified
substructures, and found three new substructures \citep{adami09}.
Despite shortage of well-relaxed cD galaxy clusters, for a better
understanding of the evolution of galaxies, the clusters at the final
evolution stage are worthwhile to be extensively investigated with
multiwavelength data. In addition, the well-relaxed clusters of
galaxies are important laboratories for studying the dark matter in
clusters. Because hot gas in the core regions of such clusters is
undisturbed by interactions with the central radio source, it is
relatively straightforward to resolve the spatial distribution of the
gravitation matter which is dominated by the dark matter
\citep{buote04}.

The regular cluster of galaxies A2589 (z=0.0414), located at
23$^{h}$24$^{m}$00$^{s}$.5, +16\degr49\arcmin29\arcsec.0
(J2000.0), with Abell richness R=0 (Abell 1958) and BM type I
\citep{bautz70}, is particularly well suited for such analysis. It
appears remarkably relaxed and does not show any obvious signs of
ongoing mergers. According to the ROSAT images, A2589 is a cluster
with a smooth X-ray morphology \citep{buote96}, and the ROSAT PSPC
observation reveals a cool gas temperature ($kT \sim 3$ keV) and a
high X-ray luminosity ($L_{\rm 0.5-2.0keV} \sim 8.5\times10^{43}$
ergs s$^{-1}$) for the central region of A2589, defined by a
radius of 0.5 Mpc \citep{david96}.  The $Chandra$ image with
higher spatial resolution shows no substantial morphological
disturbance and no cooling flow in the core of A2589
\citep{buote04}. \citet{mccarthy04} supposed an entropy injection
mechanism to explain the relaxed status of the ``non-cooling
flow'' cluster A2589. Additionally, no bright radio source is
detected at the cluster center by the NRAO VLA Sky Survey (NVSS)
\citep{bauer00}.

Previous optical studies of the galaxies in A2589 also show no
evidence supporting the presence of subclusters \citep{beers91}.
It is still a tough task to spectroscopically cover all the dwarf
galaxies with low surface brightness \citep{kambas00}. As
mentioned above, A2589 is not a rich cluster. \citet{abell58}
appraised its richness as R=0, which means less than 50 member
galaxies within the magnitude range $m_3$ to $m_3+2$ (where $m_3$
is the magnitude of the third brightest galaxy). The multicolor
optical photometry therefore becomes an alternative way for
picking up faint member galaxies in a cluster. According their
spectral energy distributions (SEDs), the redshifts of galaxies
can be estimated by the technique of photometric redshift. The
Beijing-Arizona-Taiwan-Connecticut (BATC) photometric survey is
designed for such purpose, and allows us to study the properties
of member galaxies in nearby ($z<0.1$) clusters
\citep{yuan01,yuan03,yang04,zhang10}.

\begin{figure}[!t]
\vspace{0pt} \centering
\includegraphics[width=150mm]{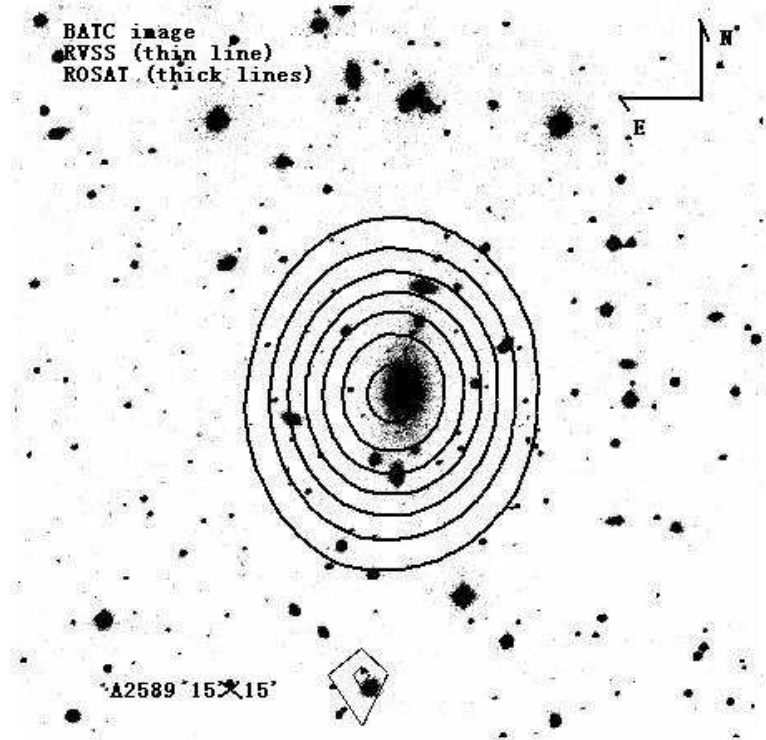}
\caption{\baselineskip 3.2mm The smoothed contours of the ROSAT
PSPC image in soft X-ray band (0.1-2.4keV) ({\sl thick line}) and
of the NVSS map at 1.4 GHz ({\sl thin line}), superimposed on
optical image of 15\arcmin $\times$ 15\arcmin in the BATC $h$
band. The sizes of gaussian smoothing windows are adopted as
30\arcsec and 2\arcmin for the radio and X-ray contours,
respectively.
  \label{fig1}}
\end{figure}

The relaxed and structureless appearance of A2589 can be well shown
by Figure~1, which presents the smoothed contours of the ROSAT
All-Sky Survey (RASS) image in soft X-ray band (0.1-2.4keV) and of
the NVSS map at 1.4 GHz, superimposed on optical image of 15\arcmin
$\times$ 15\arcmin in the BATC $h$ band. In this paper we will
present a multicolor photometric study of the relaxed cluster A2589,
using 15 intermediate filters of the BATC system which covers an
optical wavelength region from 3000 to 10000 \AA. The paper is
organized as follows. In Section 2, we describe the BATC observation
and data reduction. The SED selection of faint cluster galaxies is
given in Section 3. In Section 4, we will show the main results about
spatial distribution, dynamics, luminosity function, and star
formation properties of cluster galaxies. Finally, summary and
perspective will be given in Section 5. Throughout this paper we
adopt the $\Lambda$CDM cosmology model with H$_{0}$=70 km s$^{-1}$
Mpc$^{-1}$, $\Omega_{m}$=0.3 and $\Omega_{\Lambda}$=0.7.

\section{Observations and Data Reduction}
\subsection{BATC Observation }

The observations of A2589 were carried out with the BATC filter
system which includes 15 intermediate-band filters, namely, $a-k$,
and $m-p$, covering the whole optical wavelength range. These
intermediate filters are specifically designed to avoid most of
the known bright night-sky emission lines \citep{yan00}, and
they are mounted on the 60/90 cm f/3 Schmidt Telescope at the
Xinglong station, National Astronomical Observatories of China
(NAOC). Before October 2006, the BATC system was equipped with an
old 2048 $\times$ 2048 Ford CCD camera. The field of view was
about 1 square degree, and the spatial scale was 1.7\arcsec per
pixel. For pursuing a better spatial resolution and a higher
sensitivity in blue bands, a new E2V 4096 $\times$ 4096 CCD is now
put into service. This CCD has a quantum efficiency of 92.2\%
at 4000 \AA\ and a spatial scale of 1.35\arcsec per pixel.
The field of view is extended to 92\arcmin$\times$92\arcmin. The
pixel sizes for the old and new CCD cameras are 15 and 12 $\mu$m,
respectively, corresponding to a pixel size ratio of 5:4. The
details of the telescope, camera, and data-acquisition system can
be found elsewhere \citep{zhou01, yan00}. For
distinguishing explicitly between the SDSS and BATC filter names,
in this paper, we refer to the SDSS filters and magnitudes as {\it
u', g', r', i', z'}, with their central wavelengths of
3560, 4680, 6180, 7500 and 8870\AA. The transmissions of the BATC
and SDSS filters can be seen in \citet{yuan03}.

From September 2002 to November 2005, only 12 BATC filters, from
{\it d} to {\it p}, were taken to target A2589 with the old CCD,
discontinuously. Since 2006, the exposures in {\it a, b, c} bands
have been completed with new CCD camera. In total, we have made
169 images with more than 54 hr of exposure, which were selected
to be combined. The details of A2589 observation are given in Table 1.
\begin{table}[ht]
\centering
\begin{minipage}{125mm}
\caption[]{Parameters of the BATC Filters and the Observational
Statistics of A2589}
\end{minipage}
\footnotesize
\begin{tabular}{ccccccccc}
\hline\noalign{\smallskip}
 No. & Filter & $\lambda_{c}^a$ & FWHM &
Exposure & Seeing$^b$  & Completeness &  Indival Exposure & frames\\
   &  name & (\AA) & (\AA) & (s) & (arcsec) & magnitude & (s) & \\
\noalign{\smallskip}   \hline \noalign{\smallskip}
01 &$ a$ & 3369.4 & 277.6 & 32400  & 3.99 & 20.5 & 1200 & 27\\
02 &$ b$ & 3921.3 & 265.5 & 19200  & 3.70 & 20.5 & 1200 & 16\\
03 &$ c$ & 4205.6 & 288.5 & 13800  & 3.76 & 20.5 & 900  & 14\\
   &     &        &       &        &      &      & 600  & 2 \\
04 &$ d$ & 4528.9 & 370.6 & 12000 & 5.45  & 21.0 & 1200 & 20\\
05 &$ e$ & 4884.7 & 372.4 & 13200 & 5.01  & 21.0 & 1200 & 11\\
06 &$ f$ & 5225.1 & 333.8 & 9600  & 5.08  & 21.0 & 1200 & 8\\
07 &$ g$ & 5793.8 & 284.5 & 7200  & 4.07  & 20.0 & 1200 & 6\\
08 &$ h$ & 6093.6 & 310.4 & 6000  & 4.24  & 20.0 & 1200 & 5\\
09 &$ i$ & 6093.6 & 310.4 & 5101  &  4.41 & 19.5 & 1200 & 4 \\
   &     &        &       &        &      &      & 301  & 1\\
10 &$ j$ & 7032.3 & 163.0 & 6000  & 4.38  & 19.5 & 1200 & 5\\
11 &$ k$ & 7539.5 & 197.2 & 9000  & 4.29  & 19.0 & 1200 & 7 \\
   &     &        &       &        &      &      & 300  & 2\\
12 &$ m$ & 8012.7 & 286.6 & 12000 & 4.21  & 19.0 & 1200 & 10\\
13 &$ n$ & 8509.5 & 168.2 & 11414 & 4.35  & 19.0 & 1200 & 9\\
   &     &        &       &        &      &      & 614  & 1\\
14 &$ o$ & 9172.9 & 248.4 & 18000 & 3.97  & 18.5 & 1200 & 15\\
15 &$ p$ & 9740.7 & 272.0 & 19200 & 3.85  & 18.5 & 1200 & 16\\
\noalign{\smallskip}   \hline\noalign{\smallskip}
\end{tabular}
\parbox{130mm} {$^a$ Central wavelengths of the filters;
$^b$ This column lists the seeings of the combined images.}
\end{table}

\subsection{Data Reduction}

The bias substraction and dome flat-field correction were done on
the CCD images by using {\it Pipeline I} \citep{fan96, zhou01}
, the automatic data reduction code for the BATC
survey. The cosmic-ray and bad pixel effects were corrected by
comparing the images. Before combination, the images were
recentered, and the position calibration was performed by using
the Guide Star Catalog (GSC) \citep{jenkner90}.

For detecting sources and measuring the source fluxes within a
given aperture in the combined BATC images, {\it Pipeline II}, a
photometry package developed on the basis of DAOPHOT \citep{stetson87}
 kernel \citep{zhou03}, is taken to perform aperture
photometry. The objects with signal-to-noise ratio greater than
the threshold 3.5 $\sigma$ in $i$, $j$, and $k$ bands will be
detected. Considering the pixel size ratio of 5:4 between the old
and the new CCDs, we take an aperture radius of 4 pixels (i.e.
$r$=1.\arcsec7$\times$4=6.\arcsec8) for the images in 12 bands ($d$
to $p$), and a radius of 5 pixels (i.e.,
$r$=1\arcsec.35$\times$5=6.\arcsec8) for the images in {\it a,b,c}
bands. Flux calibration in the BATC system is performed by using
four Oke-Gunn standard stars (HD 19445, HD 84937, BD+262606 and
BD+17 4708) \citep{gunn83}, whose fluxes have been
slightly corrected by further BATC observations during photometric
nights \citep{zhou01}. To check the results of flux calibration
via standard stars, we then perform the model calibration on the
basis of the stellar SED library \citep{zhou99}. The flux
measurements derived by these two calibration methods are in
accordance with each other for most filters. As a result, the SEDs
of 3465 sources are achieved down to $i=19.5$ within our field of
view.

For assessing the measurement errors in specified magnitude, we
compared the errors using different subgroups of magnitudes with an
interval of 0.5 mag. We find that magnitude error in each filter
becomes larger at fainter depth. Errors in the $a-n$ bands are found
to be less than 0.02 mag for stars with $i < 16.5$, and more than
0.05 mag for stars with $i > 19.0$.

\section{Selection of faint member galaxies in A2589}
\subsection{Combing the BATC and SDSS SEDs}

Owing to low spatial resolution of the BATC images, it is difficult
to make star-galaxy separation via the surface brightness profiles of
detected sources.  Alternatively, the color-color diagram is a
powerful tool for morphological classification. Fortunately, A2589
has been covered by the SDSS photometric survey, and the star-galaxy
classification of all sources down to $r'=22.0$ has been performed
via their light profiles. After a cross-identification of these 3465
sources detected by the BATC photometry with the SDSS photometric
catalog, we obtain a list of 1199 galaxies brighter than $i=19.5$
within the BATC field of $58' \times 58'$. For checking the
completeness of the BATC detection of galaxies, we compare the SDSS
galaxies down to a same brightness depth (i.e., $r'<$19.5) in the
same region. Within central region with a radius of 0.5 degree, there
are 555 SDSS-detected galaxies among which 537 galaxies are detected
by the BATC, corresponding to a completeness of BATC detection of
96.8\%.
Spacial scale at cluster redshift $z=0.0414$ is 0.818 kpc/arcsec,
the typical seeing of combined images in the BATC bands is about
4.2 arcsec, corresponding to 3.4 kpc, which is smaller than the
size of a typical spiral galaxy. With the photometric catalog of
SDSS galaxies, a cross-identification of the 1199 galaxies with
i$<$19.5 is performed for estimating the  percentage of object
blending due to the seeing effect. A searching circle with a
radius of 4.3 arcsec centered at BATC galaxies is adopted, and 27
galaxies are found to have more than one counterpart within the
searching circle. The overall blending percentage is about 2.25\%.
We divide the BATC galaxies into 4 subsamples with the $i$-band
magnitude interval of 1.0 mag, ranging from 15.5 to 19.5, the
blending percentage are 2.63\%, 4.21\%, 2.08\%, 2.03\%,
respectively.

Furthermore, we take use of the NASA/IPAC Extragalactic Database
(NED) to extract available observational information of the bright
galaxies on the list. There are 81 galaxies having known
spectroscopic redshifts. Note that A2589 has not been covered by the
SDSS spectroscopic survey, the available spectroscopic redshifts of
galaxies are contributed by relevant literature \citep{bothun88,
capelato91, wegn99, smith04, hay97}. Majority of spectroscopic
redshifts were obtained by \citet{smith04}, and their selection
criteria for spectroscopy are: $R<17.0$ and $\Delta (B-R)>-0.2$. For
estimating completeness of spectroscopic sample, we apply the same
criteria to the SDSS photometric data. By using the equations in
Lupton (2005), the SDSS magnitudes can be transformed into $B$ and
$R$ magnitudes. As a result, 111 SDSS galaxies are found to satisfy
the criteria. Figure~2 shows the histograms of these 111 SDSS
galaxies and 81 galaxies with known spectroscopic redshifts. The
overall spectroscopic completeness down to $R=16.5$ should be about
$58/78\sim 74\%$. Since only 30\% of the faint galaxies with
$16.5<R<17.0$ are spectroscopically covered, the overall completeness
down to $R=17.0$ decreases to $68/108\sim 63\%$.

\begin{figure}[!t]
\vspace{0pt} \centering
\includegraphics[width=80mm]{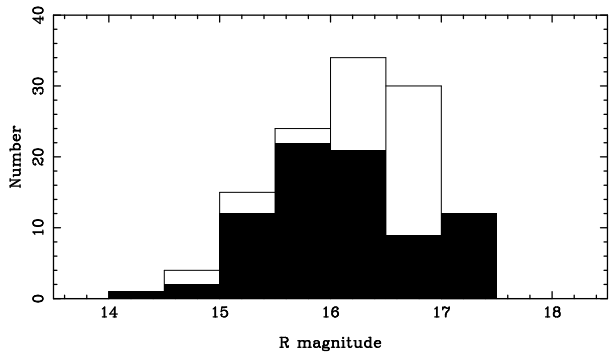}
\caption{ The histograms of the 111 SDSS galaxies selected by the
criteria: $R<17.0$ and $\Delta (B-R)>-0.2$, superposed with the
histrogram of 81 galaxies with known spectroscopic redshifts (hatched
region).
  \label{fig2}}
\end{figure}

For a given galaxy detected by the SDSS imaging observation, two
models were used to fit the observed surface brightness profile: (1)
the de Vaucouleurs model, and (2) the exponential model. For
measuring unbiased colors of galaxies, we use the dereddened SDSS
model magnitudes which are measured by their flux through equivalent
apertures in all SDSS bands. The model (exponential or de
Vaucouleurs) of higher likelihood in the r filter is chosen, and then
applied (i.e., allowing only the amplitude to vary) in the other SDSS
bands after convolving with the appropriate PSF in each band. Because
the resulting model magnitudes are calculated using the best-fit
parameters in the $r'$ band, the light is measured consistently
through the same aperture in all SDSS bands.

Figure~3(a) shows the relative SDSS and BATC SEDs of the central cD
galaxy in A2589, NGC~7647. It is clear that there is a systematic
offset between two SEDs which is called zero point. A method of
interpolation is applied to derive the zero point. Due to a larger
magnitude error in the $u'$ band, the magnitude offsets for other
four SDSS filters
are calculated by the interpolation algorithm on the basis of the
BATC magnitudes in neighboring bands, and then take the average of
the second and third largest offsets as the SED zero point.
Figure~3(b) gives the combined SED of NGC~7647 after zero point
correction.  The zero point distribution for all the galaxies is
presented in Figure~4. It can be seen that the peak of the zero point
is $-0.2$ mag, indicating that the aperture of $r=6.\arcsec8$ is
large enough to hold the majority of galaxies and to make different
seeing effect negligible. For the most extended central cD galaxy,
NGC~7647, our aperture just covers its core region, which leads to a
large zero point of nearly 2.0 mag. After zero point correction, the
combined 20-band SEDs for 2101 galaxies detected by BATC and SDSS
photometry are achieved.

\begin{figure}[ht]
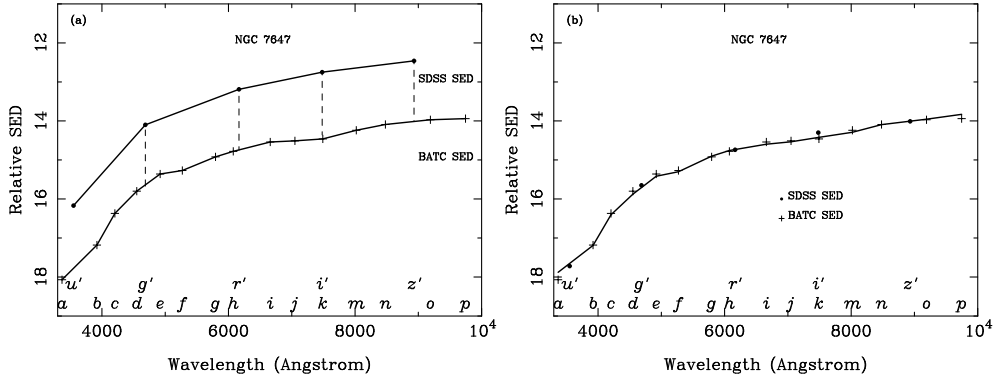

\centering
\includegraphics[width=65mm]{fig03a.eps}
\includegraphics[width=65mm]{fig03b.eps}
\caption{ \baselineskip 3.6mm Relative SEDs in the BATC and SDSS
systems for the central cD galaxy, NGC~7647, in A2589. (a) before
zero point correction (SED connections are denoted by solid lines,
and zero points in $g',r',i',z'$ bands are denoted by dashed
lines); (b): after the zero-point correction (The solid line
represents the best-fit model SED.) } \label{fig3}
\end{figure}

\begin{figure}[ht]
\centering
\includegraphics[width=75mm]{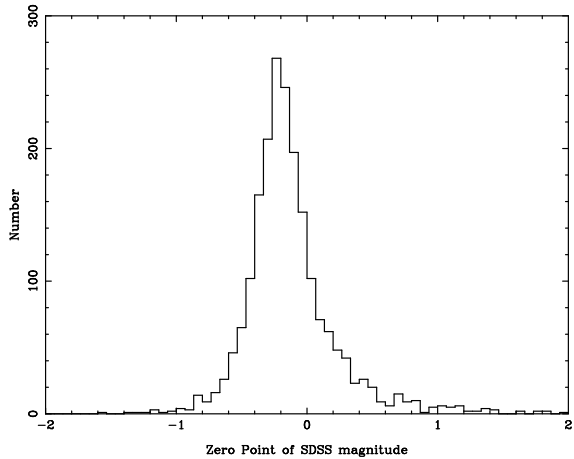}
\caption{\baselineskip 3.6mm Distribution of the zero points for
the galaxies brighter than $i=19.5$ detected by both multicolor
surveys.} \label{fig4}
\end{figure}

\subsection{Sample of spectroscopically-confirmed member galaxies}

Figure~5 shows the distribution of spectroscopic redshifts for 81
galaxies, with the bin size 0.001. The bulk of the spectroscopic
redshifts is concentrated between 0.03 and 0.06, corresponding to
A2589. In order to eliminate the background galaxies, we convert the
spectroscopic redshifts ($z_{sp}$) for the galaxies with $0.03 <
z_{sp} < 0.06$ into rest-frame velocities ($V$) by
$V=c(z_{sp}-\bar{z}_c)/(1+\bar{z}_c)$, where {\it c} is the light
speed, $\bar{z_c}$ is the cluster redshift with respect to the cosmic
background radiation. We take the NED-given cluster redshift of
$\bar{z_c}=0.0414$ for A2589 \citep{struble99}. A Gaussian fit is
then applied to the distribution of rest-frame velocities, and a
dispersion of $\sigma$=751 km s$^{-1}$ is derived. With the prevalent
selection criterium of 3$\sigma$ clipping, 68 galaxies with $-2215
<V< 2293$ \kms, corresponding to the {\it z$_{sp}$ }range from 0.034
to 0.048, are selected as member galaxies, to which we refer as
``sample I''. Table~2 lists the SDSS-given celestial coordinate in
degrees and the spectroscopic redshift in literature for these 68
member galaxies in sample I, in order by right ascension. The
embedded panel of Figure~5 shows the histogram of rest-frame
velocities for sample I with Gaussian fit.

\begin{figure}[ht]
\centering
\includegraphics[width=75mm]{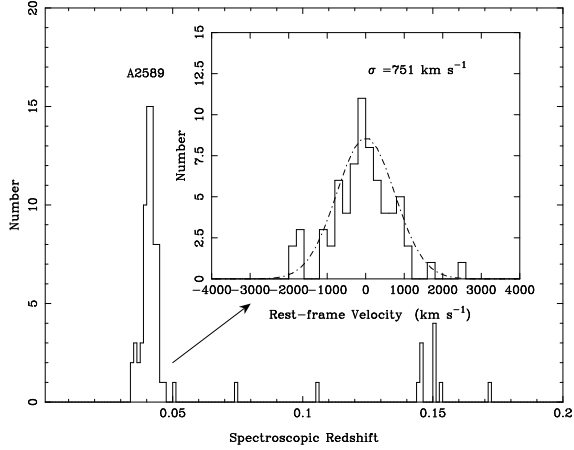}
\caption{\baselineskip 3.6mm Distribution of spectroscopic
redshifts for 81 galaxies detected by both multicolor surveys. The
embedded panel shows the histogram of rest-frame velocities for
68 member galaxies. } \label{fig5}
\end{figure}

\begin{table}[!h]
\centering\begin{minipage}{125mm}
\caption[]{Catalog of 68
Spectroscopically Confirmed Member Galaxies in A2589}
\end{minipage}
 \footnotesize
\tabcolsep 0.40mm
\begin{tabular}{rcccc|rcccc}
\hline\noalign{\smallskip}
 No. & R.A. & Decl. & $z_{\rm sp}$ &  Ref.
 & No. & R.A.& Decl & $z_{\rm sp}$ & Ref. \\
\noalign{\smallskip}   \hline \noalign{\smallskip}
    1 &350.62835693 &16.51060104 &0.037286 & (4)&  35 &350.99468994 &16.79987144 &0.040638 & (1)\\
    2 &350.64688110 &16.70728493 &0.040188 & (4)&  36 &350.99569702 &16.87491226 &0.044764 & (3)\\
    3 &350.65444946 &16.48854637 &0.037256 & (4)&  37 &350.99636841 &16.81119919 &0.042606 & (3)\\
    4 &350.65954590 &17.08135414 &0.039137 & (4)&  38 &350.99871826 &16.87028122 &0.040695 & (4)\\
    5 &350.72991943 &17.12408638 &0.039671 & (4)&  39 &351.00134277 &16.82041550 &0.034737 & (1)\\
    6 &350.74523926 &16.82660675 &0.041102 & (4)&  40 &351.00775146 &16.81094170 &0.035578 & (4)\\
    7 &350.76705933 &16.88853836 &0.039070 & (4)&  41 &351.01373291 &16.77923775 &0.041235 & (1)\\
    8 &350.79455566 &16.91345406 &0.041068 & (3)&  42 &351.02078247 &16.87013626 &0.041295 & (3)\\
    9 &350.79602051 &16.43208885 &0.044357 & (4)&  43 &351.02178955 &16.66199303 &0.040985 & (4)\\
   10 &350.83938599 &16.94395256 &0.043143 & (4)&  44 &351.02416992 &16.79190254 &0.041689 & (4)\\
   11 &350.85235596 &17.02542877 &0.039898 & (4)&  45 &351.02755737 &16.54348183 &0.040064 & (4)\\
   12 &350.87240601 &16.86256027 &0.043313 & (4)&  46 &351.03936768 &17.11158180 &0.039734 & (4)\\
   13 &350.87268066 &16.38849258 &0.043523 & (4)&  47 &351.03942871 &16.66924858 &0.044077 & (4)\\
   14 &350.88140869 &17.15295792 &0.043557 & (4)&  48 &351.04479980 &16.53371620 &0.039234 & (4)\\
   15 &350.88894653 &16.66280174 &0.039150 & (2)&  49 &351.05325317 &17.10195732 &0.044361 & (4)\\
   16 &350.90270996 &16.43026352 &0.039961 & (4)&  50 &351.06170654 &16.73438454 &0.037609 & (4)\\
   17 &350.90341187 &16.56361008 &0.044487 & (4)&  51 &351.06555176 &16.78513336 &0.041382 & (2)\\
   18 &350.92968750 &16.81964874 &0.038477 & (4)&  52 &351.07147217 &16.92798424 &0.041732 & (4)\\
   19 &350.94128418 &16.64007378 &0.040208 & (4)&  53 &351.08148193 &16.73162270 &0.044461 & (3)\\
   20 &350.94842529 &16.85229111 &0.041812 & (3)&  54 &351.08395386 &16.55175018 &0.043470 & (4)\\
   21 &350.94931030 &16.75221825 &0.035171 & (3)&  55 &351.09353638 &17.04872704 &0.041415 & (4)\\
   22 &350.95083618 &16.76874352 &0.047183 & (4)&  56 &351.09497070 &17.19940758 &0.042426 & (4)\\
   23 &350.95123291 &16.70607948 &0.042019 & (4)&  57 &351.09564209 &16.80035019 &0.042386 & (4)\\
   24 &350.96112061 &16.69865799 &0.039157 & (4)&  58 &351.13067627 &16.86814690 &0.035628 & (5)\\
   25 &350.96188354 &16.89692116 &0.040441 & (4)&  59 &351.13531494 &17.10596466 &0.040848 & (4)\\
   26 &350.96264648 &16.52305412 &0.042426 & (4)&  60 &351.14495850 &16.68141174 &0.040405 & (3)\\
   27 &350.96414185 &16.64476776 &0.034701 & (3)&  61 &351.17916870 &17.13553810 &0.041292 & (4)\\
   28 &350.96905518 &16.87223816 &0.042670 & (2)&  62 &351.19665527 &16.94142342 &0.043236 & (4)\\
   29 &350.97305298 &16.88004684 &0.044984 & (4)&  63 &351.24346924 &16.53924561 &0.042186 & (4)\\
   30 &350.97683716 &16.68067169 &0.045195 & (3)&  64 &351.30581665 &16.40709877 &0.039030 & (4)\\
   31 &350.98266602 &16.91675758 &0.041212 & (4)&  65 &351.30831909 &16.82109833 &0.043984 & (4)\\
   32 &350.98645020 &16.75003433 &0.038326 & (4)&  66 &351.33325195 &17.02133942 &0.041022 & (4)\\
   33 &350.98931885 &16.77722740 &0.041118 & (3)&  67 &351.33901978 &16.68581009 &0.041722 & (4)\\
   34 &350.99304199 &16.75839233 &0.041949 & (4)&  68 &351.39672852 &16.54359245 &0.041722 & (4)\\

\noalign{\smallskip}   \hline\noalign{\smallskip}
\end{tabular}
\parbox{125mm}
{ References: {(1) \citet{bothun88}; (2) \citet{capelato91}; (3)
\citet{wegn99}; (4) \citet{smith04}; (5) \citet{hay97}} }
\end{table}

With only 26 member galaxies in cluster A2589, \citet{capelato91}
derived a Gaussian velocity distribution with $\mu_{\rm cz}=12437 \pm
91$ \kms and $\sigma=415^{+119}_{-58}$ \kms. Based on 30 member
galaxies, \citet{beers91} characterized the distribution of measured
galaxy velocities by using the ROSTAT software. Two resistant and
robust estimators (namely, the biweight location C$_{\rm BI}$ and
scale S$_{\rm BI}$), analogous to the velocity mean and standard
deviation, are defined to characterize the velocity distribution
\citep{beers90}. \citet{beers91} found C$_{\rm
BI}$=12475$^{+199}_{-139}$ \kms and S$_{\rm BI}$=624$^{+457}_{-274}$
\kms. For 68 spectroscopically confirmed member galaxies, we use the
ROSTAT software to calculate the biweight location and scale, and
achieve C$_{\rm BI}$=12122 $\pm$ 90 \kms and S$_{\rm BI}$=737 $\pm$
85 \kms. Comparatively, the biweight location in our statistics is
smaller, and the biweight scale is larger with smaller uncertainties.
In addition, the ROSTAT statistics for the velocity distribution
shows that the number of big gaps is 0, which means the A2589 might
be a virialized system.

\subsection{Photometric Redshift Technique}

It is well known that the technique of photometric redshift can be
used to estimated the redshifts of galaxies on the basis of the
SED information given by multicolor photometric surveys. With the
development of large and deep field survey, this technique has
been widely applied \citep{lan96,
arn99, fur00}. Based on the standard
SED-fitting code called HYPERZ \citep{bol00}, the
procedures for estimating the photometric redshifts have been
developed especially for the BATC multicolor photometric system
\citep{yuan01,xia02}.
For a given object, the photometric redshift, z$_{ph}$, corresponds
to the best fit (in the $\chi^{2}$-sense) between its photometric SED
and the template SED. The SED templates for the normal galaxies are
generated by convolving the galaxy spectra in template library with
the transmission curves of the BATC and SDSS filters. In our SED
fitting, we take the templates of normal galaxies in the GISSEL98
(Galaxy Isochore Synthesis Spectral Evolution Library)
\citep{bruzual93}. The dust extinction with a reddening law of Milky
Way \citep{allen76} is adopted,  and the A$_{V}$ is flexible in a
range from 0.0 to 0.5, in steps of 0.05.
As a test, we firstly let the photometric redshift vary in a wide
range from 0.0 to 1.0, with a step of 0.01. Only five galaxies are
found to have $z_{ph}>0.5$. Then we search the $z_{ph}$ values for
1199 galaxies brighter than $i=19.5$ in a range from 0.0 to 0.5, with
a smaller step of 0.005.


During performing the HYPERZ, we develop an iterative method to
discard the bad magnitudes in the observed SED. For a given galaxy,
the first running of the HYPERZ gives the best-fit template SED, and
the observed magnitude at a specified filter with the maximum
deviation from the best-fit SED is selected. If its maximum deviation
excesses 5 times of magnitude error, we will discard this magnitude
in the observed SED, and search for the best SED fitting again. In
the worst case, three magnitudes in an observed SED are allowed to be
discarded in order to obtain a more accurate photometric redshift.

As a result, we obtain the photometric redshifts for 1199 galaxies
brighter than $i=19.5$, including the 81 galaxies with known $z_{sp}$
values. Figure~6(a) shows the redshift comparison for these 81
galaxies.  It is obvious that our {\it $z_{ph}$} estimate is
basically consistent with the {\it $z_{sp}$} values. For the 68
member galaxies, the mean photometric redshift is 0.0393, and the
standard deviation is 0.0077. The solid line denotes to {\it
$z_{ph}$}={\it $z_{sp}$}, and the dash lines indicate $\pm 3\sigma$
deviations (i.e., $3\times0.0077=0.023$) in the $z_{ph}$ estimate,
and the error bar of {\it z$_{ph}$} corresponds to 68$\%$ confidence
level in our $z_{ph}$ determination. Considering the selection
algorithm of 3$\sigma$ clipping in the $z_{ph}$ space, 66 (about
97\%) of 68 member galaxies have their $z_{ph}$ values between 0.016
and 0.062 (i.e., $\bar{z}_{ph}\pm3\sigma$), only two
spectroscopically confirmed member galaxies are missed out. This
$z_{ph}$ range (shown in Figure~6(b) with the dashed lines) is taken
as a selection criterium to select the member candidates from the
remaining 1118 ($=1199-81$) galaxies without $z_{sp}$ values.
Figure~7(a) shows the average $\vert z_{ph}-z_{sp} \vert$ as a
function of the BATC {\it i}-band magnitude. The four magnitude bins
are defined as $i<$15.5, 15.5$<i<$16.0, 16.0$<i<$16.5, and $i>$16.5.
The standard errors of $\vert z_{ph}-z_{sp} \vert$ are 0.0066,
0.0041, 0.0027, 0.0135, respectively. {\bf A similar statistics of
$z_{ph}$ uncertainties for 1199 galaxies is given in Figure~7(b),
indicating that the $z_{ph}$ uncertainty given by the SED-fitting
code HYPERZ also depends upon the magnitude. The four magnitutde bins
are defined as 15.5 $<i<$16.5, 16.5$<i<$17.5, 17.5$<i<$18.5,
18.5$<i<$19.5, and their standard errors are 0.0011, 0.0018, 0.0037,
0.0104, respectively. For the faint galaxies with greater magnitude
errors, the degree of SED fitting tends to be worse, which results in
larger uncertainties.}


\begin{figure}[!t]
\centering
\includegraphics[width=135mm]{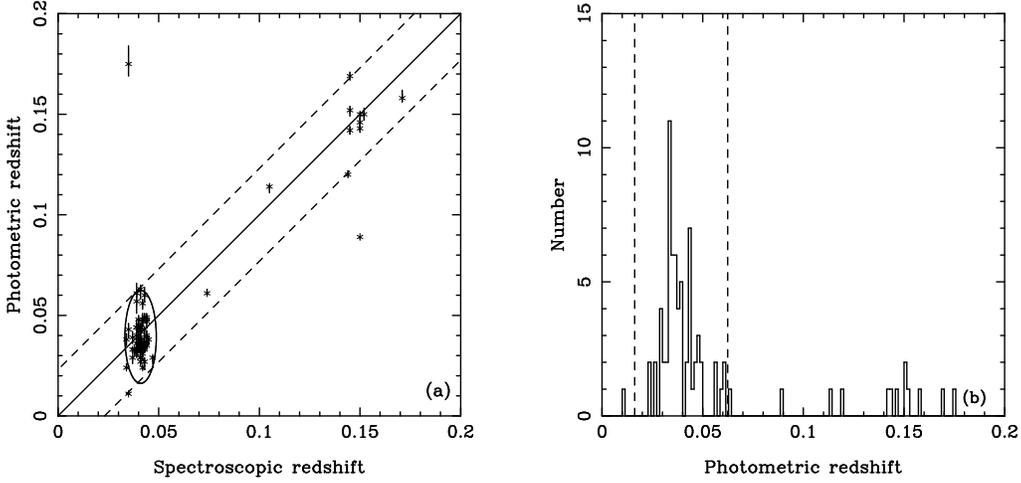}
\caption{Left: Comparison between photometric redshift ({\it
z$_{ph}$}) and spectroscopic redshift ({\it z$_{sp}$}) for 81
galaxies with known spectroscopic redshifts in the region of
A2589. Right: The distribution of  photometric redshifts of 81
known spectroscopic redshift galaxies. The dash lines are the
3$\sigma$ range of the photometric redshift.\label{fig6}}
\end{figure}

\begin{figure}[!t]
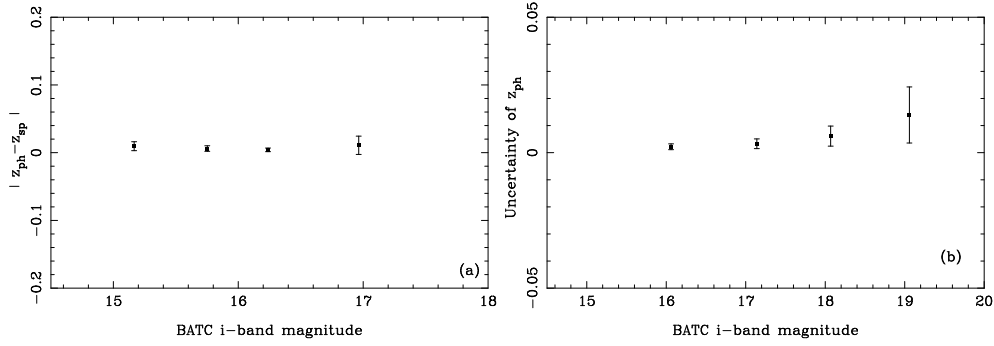

\centering
\includegraphics[width=65mm]{fig07_a.eps}
\includegraphics[width=65mm]{fig07_b.eps}
\caption{Left: Plot of the average deviations , $\vert z_{ph}-z_{sp}
\vert$, and standard errors (denoted by error bars) as a function of
BATC {\it i}-band magnitude. The four magnitude bins are defined as
$i<$15.5, 15.5$<i<$16.0, 16.0$<i<$16.5, $i>$16.5. The standard errors
of $\vert z_{ph}-z_{sp} \vert$ are 0.0066, 0.0041, 0.0027, 0.0135,
respectively. {\bf Right: Plot of the uncertainty of 1199 photometric
redshit as a function of BATC {\it i}-band magnitude. The four
magnitutde bins are defined as 15.5 $<i<$16.5, 16.5$<i<$17.5,
17.5$<i<$18.5, 18.5$<i<$19.5, and their standard errors are 0.0011,
0.0018, 0.0037,0.0104, respectively.} \label{fig7}}
\end{figure}

\begin{figure}[!h]
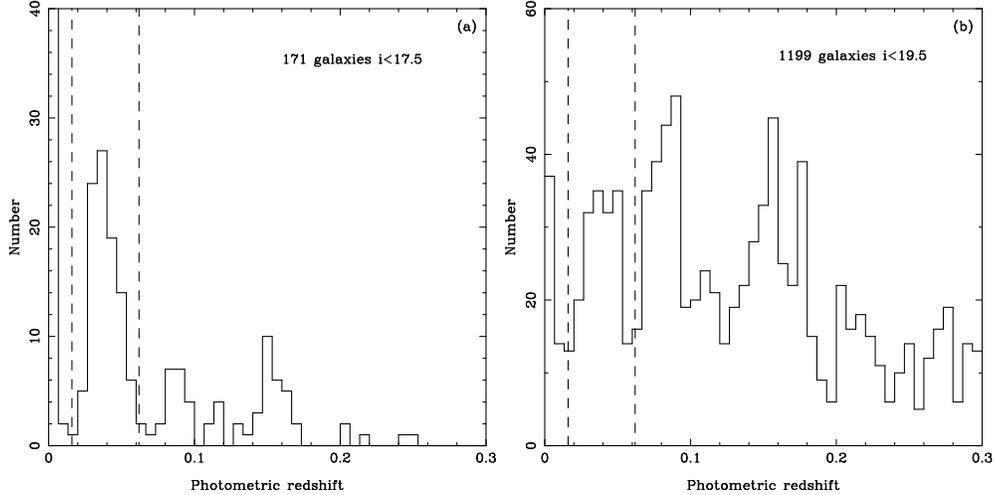

\centering
\includegraphics[width=65mm]{fig08a.eps}
\includegraphics[width=65mm]{fig08b.eps}
\caption{ \baselineskip 3.6mm Distributions of estimated photometric
redshifts for (a) 171 galaxies brighter than $i=17.5$; (b) 1199
galaxies brighter than $i=19.5$. \label{fig8}}
\end{figure}

After applying the $z_{ph}$-technique to the combined SEDs of 1118
galaxies, 110 faint galaxies (including 73 early-type galaxies and 37
late-types) are found as member galaxies candidates of A2589.
Figure~8 shows the z$_{ph}$ histograms for the galaxies in different
detection depths. Panel (a) gives the $z_{ph}$ distribution for the
171 bright galaxies with $i<17.5$, and panel (b) is for the 1199
galaxies with $i< 19.5$. The dash lines are denoted as the
photometric redshift range of cluster member candidates. At the
detection depth of $i=17.5$, corresponding to the limit of the SDSS
spectroscopy, the peak at $z_{ph}=0.04$ (contributed by cluster
A2589) is very prominent. As the magnitude limit goes to the faint
end, the peak of A2589 is still remarkable though a large number  of
background galaxies are detected, which demonstrates the reliability
of our method of combining the BATC and SDSS photometric data and the
SED-fitting procedure.

\subsection{The Color-Magnitude Correlation}

A correlation between color and absolute magnitude for early-type
galaxies, the so-called CM relation, has been found for many rich
galaxy clusters \citep[and references therein]{bower92}. For early-type
 galaxies in a cluster, the brighter
galaxies tend to have colors redder than the fainter galaxies do.
This relation can be used for verifying the membership selection of
the early-type galaxies within the BATC field \citep{yuan01}. In
order to select the member early-type galaxies, we take the Hubble
types of the best-fit SED templates as morphology classifications.
Figure~9 presents the correlation between the color index $u'-p$
versus $h$-magnitude for the known and new early-type member
candidates. The open circles denote newly selected early-types and
the filled circles denote the known early-types. The solid line
denotes the least-square fit for the linear relation with 65 known
early-type galaxies:$ u'-p=-0.26(\pm0.04)h+7.53(\pm0.70)$. The dashed
lines represent $\pm 1 \sigma$ deviation. Taking all photometric
early-types into account, an horizontal line, $u'-p=3.15$, seems to
be also a good model of the red sequence, which means red galaxies in
A2589 might have a typical $u'-p$ color. {\bf It is noticed that the good
CM relation might be artificial because red galaxies are likely to be
classified as early-types by the HYPERZ code.}

As shown in Figure~9, most of early-type candidates agree with the
CM correlation derived by the known early-type galaxies. There are
8 early-type candidates with colors beyond the $2 \sigma$
deviation of intercept that have probably been misclassified.
Taking only the templates of late-type galaxies, we apply the
SED-fitting procedures again on these 8 candidates. As a result, 4
of them are found to have the best-fit template SEDs of Sa galaxies
with redshifts between 0.016 and 0.062, and they are therefore
regarded as late-type member galaxies. The remaining 4 candidates
are excluded from the list of member galaxies.

\begin{figure}[!b] \centering
\includegraphics[width=75mm]{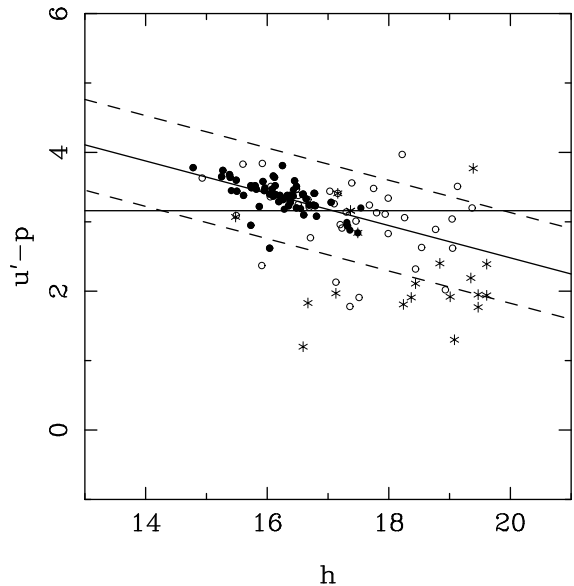}
\caption{Color-magnitude relation for 130 early-type galaxies in
A2589, including 65 early-type galaxies with known spectroscopic
redshifts ({\it filled circles}) and 65 newly-selected member
candidates ({\it open circles}). The 44 late-type member galaxies are
also plotted ({\it stars}). A linear fit is given to the 65
spectroscopically-confirmed early-type members, and the dash lines
correspond to the $\pm1\sigma$ of the linear fitting. \label{fig9}}
\end{figure}

Finally, we obtain a list of 106 ($=110-4$) newly-selected member
galaxies. Combined with the 68 spectroscopically confirmed member
galaxies, we have obtained an enlarged sample of 174 member galaxies
which includes 130 early-type galaxies and 44 late-type galaxies. We
refer to this sample as ``sample II'' hereafter. Table~3 presents the
catalog of SED information for the 106 newly-selected members,
including the ID number, the SDSS-given celestial coordinate in
degrees, photometric redshift, morphological class of the best-fit
template, and combined SED information. The classification indices in
a range from 1 to 7 are defined to denote E, S0, Sa, Sb, Sc, Sd and
Im galaxies, respectively. The magnitude of 99.00 means non-detection
in the specified band.

\begin{table}[t]
\centering \begin{minipage}{125mm}

\caption{Catalog of 106 Newly-selected Candidates of Member
Galaxies in A2589}\end{minipage}

\scriptsize \tabcolsep 0.35mm
\begin{tabular}{rcccccccccccccccccccccccc}
\hline\noalign{\smallskip} {No.}&{R.A.} &{Decl.} & {$z_{\rm ph}$} &
{$T$} & {$a$} & {$b$} & {$c$} & {$d$} & {$e$} & {$f$} & {$g$} &
{$h$} & {$i$} & {$j$} & {$k$} & {$m$}
& {$n$} & {$o$} & {$p$} & {$u'$} & {$g'$} &{$r'$} &{$i'$} &{$z'$}\\
\noalign{\smallskip}\hline\noalign{\smallskip}
    1& 350.52581787& 16.92402267&  0.048& 1& 19.97& 20.47& 19.79& 19.14& 18.77& 18.73& 18.37& 18.22& 18.02& 17.91& 99.00& 17.92& 17.96& 17.65& 17.31& 21.28& 19.12& 17.71& 17.96& 17.71\\
    2& 350.53906250& 16.87807274&  0.056& 2& 18.81& 18.14& 17.90& 17.46& 17.11& 16.89& 16.64& 16.51& 16.29& 16.21& 16.21& 16.02& 15.90& 15.83& 15.65& 18.90& 17.37& 16.55& 16.12& 15.81\\
    3& 350.53964233& 16.80180168&  0.025& 7& 19.67& 18.74& 18.76& 18.87& 18.76& 18.78& 18.73& 18.59& 18.59& 18.46& 18.61& 18.57& 18.41& 18.95& 18.30& 20.28& 18.78& 18.83& 18.64& 17.51\\
    4& 350.54162598& 16.67333031&  0.033& 3& 17.46& 17.12& 16.89& 16.80& 16.70& 16.64& 16.63& 16.59& 16.27& 16.29& 16.56& 16.42& 16.43& 16.32& 16.44& 17.64& 16.83& 16.59& 16.38& 16.31\\
    5& 350.54580688& 17.26466942&  0.044& 1& 18.43& 18.17& 17.35& 16.98& 16.61& 16.48& 16.12& 16.05& 15.82& 15.84& 15.83& 15.58& 15.48& 15.41& 15.29& 18.65& 16.88& 16.08& 15.71& 15.40\\
    6& 350.55383301& 16.42631340&  0.030& 4& 99.00& 17.89& 17.36& 17.30& 17.17& 17.09& 17.01& 16.94& 16.64& 16.68& 16.78& 16.67& 16.62& 16.51& 16.42& 18.28& 17.38& 16.91& 16.72& 16.53\\
    7& 350.56356812& 16.40281296&  0.057& 1& 99.00& 19.14& 18.76& 18.34& 17.92& 17.67& 17.50& 17.38& 17.12& 17.11& 17.02& 16.83& 16.74& 16.65& 16.59& 20.31& 18.26& 17.32& 16.91& 16.70\\
    8& 350.58166504& 16.80896950&  0.046& 3& 99.00& 19.30& 19.08& 19.27& 19.09& 19.00& 18.84& 18.76& 18.47& 18.42& 18.83& 18.39& 18.20& 17.90& 18.04& 27.70& 19.21& 18.70& 18.44& 18.47\\
    9& 350.59280396& 16.95937729&  0.027& 2& 21.58& 20.38& 20.05& 20.26& 19.60& 19.54& 19.30& 19.37& 19.17& 19.00& 19.23& 19.09& 18.62& 18.73& 18.58& 21.78& 19.96& 19.39& 19.01& 18.84\\
   10& 350.59347534& 17.01498795&  0.061& 1& 18.23& 18.19& 17.76& 17.08& 16.59& 16.36& 16.04& 15.92& 15.67& 15.61& 15.56& 15.34& 15.28& 15.19& 15.02& 18.86& 16.91& 15.98& 15.53& 15.21\\
   11& 350.59823608& 17.03547859&  0.060& 4& 18.41& 18.12& 17.76& 17.67& 17.47& 17.34& 17.20& 17.13& 17.01& 16.74& 17.01& 16.81& 16.77& 16.76& 16.68& 18.65& 17.63& 17.17& 16.84& 16.73\\
   12& 350.60241699& 17.10379982&  0.049& 1& 19.14& 18.44& 17.81& 17.45& 17.06& 16.95& 16.61& 16.51& 16.30& 16.28& 16.26& 16.04& 15.95& 15.88& 15.78& 19.16& 17.34& 16.55& 16.17& 15.87\\
   13& 350.61746216& 16.42463112&  0.023& 1& 99.00& 20.87& 23.37& 21.43& 20.31& 20.35& 19.81& 19.77& 19.44& 19.95& 19.31& 19.61& 19.63& 18.89& 19.21& 21.89& 20.76& 19.95& 19.44& 19.11\\
   14& 350.63101196& 16.37726974&  0.048& 1& 99.00& 20.15& 20.43& 18.97& 19.07& 19.17& 18.84& 18.77& 18.34& 18.68& 99.00& 18.76& 18.60& 18.93& 18.08& 21.33& 20.02& 18.94& 18.48& 18.07\\
   15& 350.63424683& 16.78801537&  0.017& 1& 19.89& 19.24& 20.16& 19.33& 18.95& 18.87& 18.54& 18.42& 18.16& 18.04& 17.96& 17.83& 17.70& 17.72& 17.48& 20.93& 19.24& 18.43& 17.92& 17.62\\
   16& 350.63787842& 16.44768906&  0.034& 1& 99.00& 18.21& 18.07& 17.93& 17.80& 17.77& 17.71& 17.68& 17.49& 17.48& 17.61& 17.59& 17.45& 17.26& 17.38& 19.08& 17.91& 17.62& 17.52& 17.63\\
   17& 350.66183472& 16.35647392&  0.053& 1& 99.00& 18.22& 19.08& 18.75& 18.23& 18.05& 17.84& 17.77& 17.48& 17.38& 17.45& 17.31& 17.12& 17.09& 17.09& 20.48& 18.57& 17.74& 17.35& 17.08\\
   18& 350.66513062& 16.86167526&  0.024& 3& 17.74& 17.16& 16.55& 16.31& 16.00& 15.89& 15.59& 15.48& 15.21& 15.20& 15.11& 15.00& 14.85& 14.73& 14.68& 17.75& 16.24& 15.47& 15.06& 14.74\\
   19& 350.68112183& 16.36630630&  0.037& 1& 99.00& 18.48& 18.64& 18.42& 18.31& 18.25& 18.14& 18.03& 17.87& 17.76& 18.11& 17.87& 17.75& 17.87& 17.70& 19.81& 18.33& 18.05& 17.81& 17.88\\
   20& 350.74844360& 17.16712379&  0.043& 1& 18.98& 18.49& 17.84& 17.45& 17.04& 16.91& 16.57& 16.48& 16.28& 16.25& 16.19& 15.97& 15.91& 15.79& 15.74& 19.21& 17.31& 16.51& 16.12& 15.82\\
   21& 350.75390625& 16.79937553&  0.024& 2& 19.40& 17.97& 17.70& 17.55& 17.37& 17.32& 17.22& 17.13& 16.90& 16.86& 16.93& 16.86& 16.79& 16.66& 16.61& 18.74& 17.49& 17.09& 16.85& 16.73\\
   22& 350.75997925& 17.14521217&  0.039& 3& 19.86& 19.90& 19.73& 20.21& 19.71& 19.59& 19.28& 19.20& 18.92& 19.17& 18.94& 18.87& 18.52& 18.41& 18.72& 21.14& 19.85& 19.22& 18.91& 18.76\\
   23& 350.76577759& 16.54315758&  0.049& 1& 20.30& 20.17& 20.00& 19.01& 18.83& 18.77& 18.41& 18.44& 18.28& 18.27& 18.29& 18.16& 18.24& 18.04& 18.40& 20.72& 19.24& 18.01& 18.31& 18.09\\
   24& 350.77868652& 17.10724640&  0.024& 1& 20.60& 19.35& 19.24& 19.50& 19.68& 19.58& 19.74& 19.38& 19.15& 19.26& 19.36& 19.24& 18.96& 18.83& 18.27& 20.94& 19.65& 19.26& 19.09& 19.02\\
   25& 350.78274536& 16.53175163&  0.044& 1& 18.69& 17.89& 16.86& 15.96& 15.78& 15.60& 15.09& 14.93& 14.69& 14.55& 14.42& 14.30& 14.22& 14.11& 14.08& 17.71& 15.79& 14.75& 14.54& 15.16\\
   26& 350.79830933& 16.60096359&  0.043& 1& 21.28& 20.15& 19.27& 19.26& 19.10& 19.09& 18.99& 18.93& 18.85& 18.82& 19.27& 18.91& 18.58& 18.54& 19.00& 21.02& 19.26& 18.85& 18.62& 18.75\\
   27& 350.80563354& 16.58921432&  0.031& 1& 18.68& 18.03& 17.76& 17.80& 17.64& 17.56& 17.55& 17.48& 17.24& 17.24& 17.33& 17.26& 17.14& 17.26& 17.15& 18.95& 17.79& 17.48& 17.28& 17.13\\
   28& 350.82449341& 16.66783524&  0.049& 1& 20.62& 19.83& 19.37& 19.16& 18.85& 18.71& 99.00& 18.54& 18.33& 18.26& 18.43& 18.27& 18.22& 18.15& 17.70& 20.33& 19.10& 18.59& 18.35& 18.10\\
   29& 350.82653809& 17.22652817&  0.039& 2& 18.35& 18.39& 17.60& 17.49& 17.24& 17.10& 16.78& 16.71& 16.49& 16.47& 16.45& 16.25& 16.18& 16.08& 16.02& 18.79& 17.39& 16.71& 16.34& 16.12\\
   30& 350.82934570& 16.66695786&  0.042& 1& 99.00& 19.18& 22.12& 21.04& 20.26& 19.84& 19.63& 19.42& 19.39& 19.38& 19.75& 19.36& 18.77& 18.94& 18.26& 21.60& 20.64& 19.74& 19.32& 18.99\\
   31& 350.83618164& 16.79516220&  0.044& 2& 18.94& 18.48& 18.03& 17.98& 17.76& 17.69& 17.57& 17.51& 17.31& 17.34& 17.34& 17.24& 17.16& 17.15& 17.20& 19.11& 17.95& 17.53& 17.28& 17.10\\
   32& 350.85662842& 16.77861404&  0.026& 4& 19.63& 19.39& 19.17& 19.20& 19.10& 19.08& 18.82& 18.77& 18.67& 18.67& 18.45& 18.57& 18.13& 18.28& 18.15& 20.28& 19.18& 18.71& 18.46& 18.30\\
   33& 350.87145996& 16.61509132&  0.030& 1& 99.00& 21.13& 21.10& 19.81& 19.79& 19.72& 19.44& 19.09& 18.92& 19.01& 19.11& 19.02& 18.78& 18.99& 18.28& 21.46& 19.95& 19.30& 18.94& 18.74\\
   34& 350.87319946& 16.90993500&  0.048& 5& 18.74& 17.81& 17.28& 17.19& 17.00& 16.93& 16.74& 16.67& 16.38& 16.39& 16.55& 16.40& 16.33& 16.26& 16.18& 18.01& 17.12& 16.71& 16.53& 16.03\\
   35& 350.87722778& 16.76000595&  0.031& 1& 20.83& 19.47& 18.77& 18.22& 17.91& 17.83& 17.53& 17.46& 17.20& 17.12& 17.13& 17.00& 16.84& 16.76& 16.81& 19.82& 18.19& 17.42& 17.07& 16.79\\
   36& 350.88580322& 17.04202461&  0.059& 3& 99.00& 20.15& 19.62& 19.63& 19.33& 19.19& 18.86& 18.66& 18.42& 18.42& 18.61& 18.18& 18.24& 18.16& 18.08& 21.35& 19.19& 18.80& 18.40& 22.37\\
   37& 350.89169312& 16.81429672&  0.050& 3& 20.64& 99.00& 19.21& 19.50& 19.07& 18.98& 18.75& 18.59& 18.03& 18.36& 18.32& 18.21& 18.01& 18.14& 17.76& 20.76& 19.36& 18.63& 18.28& 18.07\\
   38& 350.89550781& 16.77338600&  0.023& 1& 19.77& 19.65& 18.79& 18.77& 18.44& 18.39& 18.15& 17.99& 17.82& 17.80& 17.67& 17.67& 17.49& 17.35& 17.53& 20.36& 18.68& 17.97& 17.63& 17.39\\
   39& 350.90295410& 16.65809059&  0.034& 1& 19.60& 18.60& 18.01& 17.57& 17.16& 17.07& 16.79& 16.69& 16.48& 16.47& 16.41& 16.25& 16.10& 16.07& 15.95& 19.16& 17.46& 16.70& 16.34& 16.04\\
   40& 350.90997314& 17.03512764&  0.044& 4& 20.55& 19.23& 20.50& 20.07& 19.66& 19.58& 19.49& 19.47& 19.30& 19.37& 19.41& 19.05& 18.89& 18.43& 19.26& 21.03& 19.95& 19.41& 19.07& 18.88\\
   41& 350.91564941& 17.15081024&  0.024& 4& 99.00& 19.85& 19.81& 19.91& 19.96& 19.89& 19.78& 19.61& 19.29& 19.39& 19.48& 19.46& 19.15& 19.57& 18.51& 20.45& 20.02& 19.47& 19.14& 20.33\\
   42& 350.92758179& 16.63711929&  0.038& 1& 99.00& 21.10& 19.45& 18.70& 18.50& 18.36& 18.15& 17.99& 17.71& 17.69& 17.66& 17.54& 17.33& 17.31& 17.10& 20.44& 18.71& 17.91& 17.54& 17.35\\
   43& 350.93566895& 16.57441330&  0.056& 1& 19.92& 19.62& 19.04& 18.54& 18.12& 18.04& 17.80& 17.68& 17.42& 17.34& 17.40& 17.20& 17.10& 17.16& 16.93& 20.17& 18.46& 17.69& 17.34& 17.03\\
   44& 350.94158936& 16.53168869&  0.018& 1& 99.00& 20.07& 20.71& 21.04& 19.87& 20.02& 19.49& 19.44& 18.90& 19.61& 18.81& 18.88& 18.96& 18.52& 18.77& 21.69& 20.28& 19.41& 18.99& 18.63\\
   45& 350.94189453& 16.83019829&  0.043& 1& 19.53& 18.82& 18.11& 17.73& 17.31& 17.21& 16.91& 16.78& 16.57& 16.51& 16.53& 16.30& 16.17& 16.04& 16.02& 19.43& 17.59& 16.75& 16.35& 16.08\\
   46& 350.94229126& 16.98780251&  0.028& 4& 19.98& 19.84& 20.99& 19.90& 19.57& 19.54& 19.20& 19.47& 19.32& 19.30& 19.21& 19.18& 19.03& 18.43& 19.84& 20.88& 19.91& 19.34& 18.95& 18.75\\
   47& 350.95550537& 17.13400650&  0.040& 1& 20.20& 19.64& 19.98& 19.63& 19.49& 19.50& 19.22& 19.04& 18.86& 18.90& 18.90& 18.62& 18.50& 18.74& 18.44& 21.48& 19.73& 19.14& 18.78& 18.45\\
   48& 350.95840454& 16.35474205&  0.053& 1& 20.36& 20.85& 21.16& 20.08& 19.73& 19.67& 19.17& 19.66& 19.24& 18.83& 19.62& 19.04& 18.93& 19.11& 18.84& 21.15& 20.14& 19.62& 19.16& 19.02\\
   49& 350.96127319& 16.79039574&  0.053& 2& 19.94& 21.01& 19.55& 21.15& 20.47& 20.73& 20.44& 20.40& 19.45& 19.56& 19.72& 19.51& 19.46& 19.79& 19.08& 22.11& 21.00& 20.14& 19.80& 19.44\\
   50& 350.96603394& 16.67181778&  0.045& 1& 20.55& 19.06& 18.82& 18.18& 17.82& 17.72& 17.45& 17.30& 17.11& 17.09& 17.03& 16.85& 16.75& 16.66& 16.79& 19.93& 18.12& 17.32& 16.95& 16.64\\
   51& 350.96640015& 17.02894592&  0.026& 3& 19.82& 19.08& 19.00& 18.88& 18.63& 18.58& 18.36& 18.37& 18.24& 18.14& 18.20& 18.16& 18.15& 17.83& 18.14& 20.05& 18.76& 18.53& 18.06& 17.97\\
   52& 350.96762085& 16.92559433&  0.053& 1& 22.61& 19.19& 18.74& 18.29& 17.92& 17.81& 17.47& 17.39& 17.20& 17.23& 17.07& 16.92& 16.85& 16.77& 16.68& 20.24& 18.25& 17.09& 17.09& 16.80\\
   53& 350.97122192& 16.77782631&  0.046& 1& 22.02& 99.00& 18.92& 18.75& 18.30& 18.23& 17.96& 17.80& 17.69& 17.48& 17.53& 17.41& 17.28& 17.15& 17.14& 20.27& 18.62& 17.81& 17.42& 17.17\\
   54& 350.97134399& 16.73224449&  0.036& 3& 20.68& 99.00& 20.16& 19.77& 19.69& 19.56& 19.49& 19.35& 19.41& 18.94& 19.12& 18.96& 18.85& 18.66& 18.91& 21.10& 19.82& 19.28& 18.95& 18.81\\
   55& 350.97305298& 16.64459801&  0.045& 3& 22.17& 99.00& 20.97& 20.47& 19.80& 20.08& 20.28& 19.58& 19.32& 19.46& 20.45& 19.37& 19.84& 18.82& 18.59& 23.03& 20.17& 19.60& 19.32& 19.23\\
   56& 350.97525024& 16.46105576&  0.053& 1& 99.00& 20.26& 20.66& 19.56& 19.05& 18.75& 18.52& 18.39& 18.17& 18.07& 18.15& 17.90& 17.71& 17.54& 17.92& 21.26& 19.35& 18.37& 17.96& 17.63\\
   57& 350.97549438& 16.53599358&  0.046& 1& 99.00& 20.10& 19.03& 18.94& 18.59& 18.47& 18.19& 18.08& 17.90& 17.80& 17.79& 17.66& 17.53& 17.48& 17.54& 20.27& 18.81& 18.10& 17.75& 17.50\\
   58& 350.99014282& 16.72022820&  0.052& 1& 21.11& 20.54& 20.19& 20.41& 20.04& 19.83& 19.66& 19.26& 19.25& 19.02& 19.64& 18.89& 18.71& 19.04& 18.41& 21.44& 20.16& 19.42& 19.12& 19.03\\
   59& 351.00198364& 16.89023781&  0.060& 2& 19.89& 19.15& 20.41& 19.69& 19.18& 19.08& 18.74& 18.83& 18.55& 18.87& 18.73& 18.31& 18.27& 18.25& 18.29& 21.27& 19.50& 18.81& 18.47& 18.25\\
   60& 351.00668335& 17.24337769&  0.046& 7& 19.61& 19.78& 19.50& 19.50& 19.58& 19.54& 19.48& 19.29& 19.32& 18.99& 19.80& 19.10& 19.53& 18.56& 18.72& 20.64& 19.50& 19.32& 19.09& 19.51\\
   \noalign{\smallskip}\hline
\end{tabular}
\end{table}

\begin{table}
\setcounter{table}{2}

 \centering

\begin{minipage}{35mm}
 \caption{\it --- Continued.}\end{minipage}\vspace{0pt}

\small
\scriptsize \tabcolsep 0.35mm
\begin{tabular}{clccccccccccccccccccccccc}
\hline\noalign{\smallskip} {No.}&{R.A.} &{Decl.} & {$z_{\rm ph}$}
& {$T$} & {$a$} & {$b$} & {$c$} & {$d$} & {$e$} & {$f$} & {$g$} &
{$h$} & {$i$} & {$j$} & {$k$} & {$m$}
& {$n$} & {$o$} & {$p$} & {$u'$} & {$g'$} &{$r'$} &{$i'$} &{$z'$}\\
\noalign{\smallskip}\hline\noalign{\smallskip}
   61& 351.00845337& 16.75835037&  0.021& 1& 20.54& 99.00& 18.60& 18.87& 18.46& 18.36& 18.09& 17.94& 17.76& 17.75& 17.61& 17.54& 17.40& 17.24& 17.35& 20.46& 18.73& 17.96& 17.61& 17.29\\
   62& 351.01538086& 16.60781860&  0.058& 1& 21.07& 19.52& 18.94& 18.65& 18.25& 18.12& 17.87& 17.75& 17.54& 17.45& 17.46& 17.25& 17.21& 17.11& 17.00& 20.48& 18.61& 17.45& 17.47& 17.14\\
   63& 351.01727295& 16.77843475&  0.024& 3& 20.25& 19.79& 99.00& 19.56& 19.27& 19.35& 19.02& 18.93& 18.68& 18.71& 18.65& 18.68& 18.43& 18.67& 18.49& 25.43& 19.59& 18.92& 18.59& 18.56\\
   64& 351.02890015& 16.76558304&  0.020& 1& 20.00& 22.55& 20.23& 19.93& 19.76& 19.77& 19.47& 19.31& 18.71& 19.26& 18.88& 18.92& 18.80& 19.19& 18.37& 21.94& 20.02& 19.23& 18.91& 18.68\\
   65& 351.03451538& 16.70536232&  0.035& 3& 99.00& 18.89& 20.16& 20.05& 19.87& 19.88& 19.71& 19.35& 19.21& 19.51& 19.07& 19.08& 18.55& 18.87& 18.28& 21.10& 20.01& 19.31& 19.08& 18.77\\
   66& 351.03710938& 16.74093056&  0.023& 2& 23.21& 19.34& 19.10& 19.17& 18.77& 18.71& 18.39& 18.26& 18.15& 18.14& 18.04& 17.89& 17.85& 17.67& 17.53& 20.59& 19.08& 18.37& 17.91& 17.68\\
   67& 351.05773926& 17.07377625&  0.042& 3& 20.37& 99.00& 19.75& 20.35& 19.93& 20.02& 20.09& 19.51& 19.24& 19.22& 20.10& 19.33& 18.87& 19.21& 17.99& 21.51& 20.19& 19.51& 19.16& 19.09\\
   68& 351.05862427& 16.73395157&  0.039& 2& 20.61& 19.17& 18.44& 18.03& 17.71& 17.64& 17.36& 17.23& 17.07& 17.08& 16.98& 16.86& 16.64& 16.56& 16.57& 19.48& 17.91& 17.20& 16.93& 16.70\\
   69& 351.06436157& 16.63640594&  0.030& 4& 19.44& 18.88& 19.31& 18.93& 18.71& 18.66& 18.38& 18.44& 18.26& 18.23& 18.29& 18.08& 18.03& 17.76& 17.92& 20.03& 18.89& 18.43& 18.09& 17.84\\
   70& 351.07086182& 16.99158096&  0.024& 1& 99.00& 20.32& 19.20& 19.63& 19.40& 19.30& 18.90& 18.90& 18.77& 18.77& 18.69& 18.66& 18.44& 18.34& 18.28& 21.00& 19.61& 18.89& 18.59& 18.36\\
   71& 351.07095337& 16.57386208&  0.044& 1& 19.72& 19.19& 18.01& 17.58& 17.19& 17.06& 16.79& 16.67& 16.47& 16.43& 16.38& 16.22& 16.10& 15.99& 15.91& 19.23& 17.46& 16.68& 16.28& 16.00\\
   72& 351.07333374& 17.14213371&  0.049& 4& 19.34& 18.67& 18.68& 18.73& 18.49& 18.29& 18.31& 18.24& 17.95& 17.77& 18.06& 17.91& 17.85& 17.98& 17.91& 19.72& 18.69& 18.30& 17.97& 17.78\\
   73& 351.08377075& 16.50154114&  0.048& 1& 19.22& 19.18& 19.89& 19.37& 19.31& 19.35& 19.23& 19.10& 18.81& 18.54& 19.12& 18.79& 18.97& 19.08& 19.04& 20.57& 19.52& 19.15& 18.94& 18.91\\
   74& 351.08947754& 17.21441841&  0.017& 3& 20.79& 21.03& 20.33& 20.30& 19.88& 20.04& 19.59& 19.61& 19.30& 19.23& 19.33& 19.16& 19.05& 19.37& 19.14& 21.53& 20.25& 19.62& 19.25& 18.87\\
   75& 351.09808350& 16.88235664&  0.057& 1& 19.18& 19.37& 19.29& 18.75& 18.43& 18.26& 18.02& 17.90& 17.71& 17.73& 17.64& 17.46& 17.35& 17.42& 17.32& 20.40& 18.63& 17.87& 17.71& 17.23\\
   76& 351.11566162& 16.69385719&  0.047& 1& 19.95& 19.04& 18.35& 18.00& 17.59& 17.47& 17.23& 17.10& 16.87& 16.87& 16.80& 16.62& 16.48& 16.46& 16.35& 19.61& 17.89& 17.10& 16.71& 16.43\\
   77& 351.11795044& 16.47721481&  0.023& 3& 19.64& 21.00& 20.46& 20.28& 19.91& 20.04& 19.73& 19.37& 19.21& 19.47& 19.24& 19.00& 18.83& 18.61& 19.61& 21.22& 20.05& 19.43& 19.09& 18.86\\
   78& 351.12063599& 16.41191292&  0.060& 3& 19.79& 19.44& 19.54& 19.19& 18.93& 18.75& 18.57& 18.52& 18.25& 18.18& 18.31& 18.07& 17.85& 17.89& 17.78& 20.45& 19.19& 18.42& 18.19& 17.93\\
   79& 351.12286377& 17.11305428&  0.048& 3& 99.00& 19.32& 20.58& 19.98& 19.77& 19.77& 19.66& 19.32& 19.38& 19.04& 19.26& 18.92& 18.64& 19.10& 18.72& 21.28& 20.00& 19.36& 19.06& 18.91\\
   80& 351.12655640& 16.86726189&  0.050& 3& 19.86& 18.40& 17.88& 17.72& 17.60& 17.59& 17.54& 17.48& 17.17& 17.26& 17.34& 17.25& 17.16& 17.31& 17.31& 17.09& 17.57& 17.53& 17.53& 16.89\\
   81& 351.13357544& 16.66821671&  0.042& 1& 19.57& 19.08& 18.33& 17.97& 17.67& 17.60& 17.34& 17.20& 17.01& 16.95& 16.95& 16.77& 16.67& 16.59& 16.54& 19.50& 17.92& 17.22& 16.82& 16.57\\
   82& 351.16860962& 16.48097610&  0.056& 4& 19.08& 19.39& 18.77& 18.76& 18.49& 18.43& 18.27& 18.16& 17.88& 17.88& 17.93& 17.67& 17.71& 17.54& 17.48& 19.97& 18.69& 18.12& 17.81& 17.60\\
   83& 351.17163086& 16.57984734&  0.048& 3& 21.16& 20.21& 20.15& 19.50& 19.25& 19.19& 18.87& 18.84& 18.74& 18.41& 18.43& 18.32& 18.20& 18.11& 18.37& 20.77& 19.48& 18.74& 18.41& 18.16\\
   84& 351.21090698& 17.24355888&  0.060& 2& 18.82& 20.16& 19.54& 19.82& 19.45& 19.15& 18.99& 19.05& 18.80& 18.59& 18.87& 18.61& 18.81& 18.13& 18.34& 20.96& 19.72& 18.91& 18.79& 18.52\\
   85& 351.21206665& 16.90178680&  0.048& 1& 21.39& 19.59& 19.38& 19.17& 18.73& 18.62& 18.33& 18.22& 18.03& 17.91& 17.89& 17.79& 17.67& 17.55& 17.29& 20.60& 19.01& 18.05& 17.89& 17.62\\
   86& 351.21966553& 16.53191757&  0.059& 3& 19.22& 20.53& 19.63& 19.52& 19.03& 19.04& 19.05& 18.67& 18.57& 18.56& 18.40& 18.33& 18.17& 18.28& 18.47& 24.55& 19.46& 18.71& 18.36& 18.12\\
   87& 351.22222900& 16.40045166&  0.017& 1& 19.30& 99.00& 19.30& 19.94& 19.47& 19.41& 19.31& 19.02& 18.90& 18.56& 18.72& 18.73& 18.60& 18.16& 18.12& 22.09& 19.75& 19.01& 18.69& 18.38\\
   88& 351.24563599& 16.71162224&  0.047& 3& 20.25& 19.46& 18.89& 18.53& 18.23& 18.14& 17.91& 17.86& 17.61& 17.62& 17.55& 17.47& 17.40& 17.09& 17.14& 19.66& 18.42& 17.82& 17.50& 17.35\\
   89& 351.24801636& 17.16534615&  0.024& 1& 23.67& 20.18& 19.45& 19.96& 19.62& 19.56& 19.20& 19.13& 18.91& 18.95& 18.70& 18.55& 18.52& 18.26& 17.96& 21.47& 19.93& 19.07& 18.62& 18.38\\
   90& 351.26184082& 16.57231903&  0.023& 3& 18.84& 20.03& 20.31& 20.80& 20.38& 20.48& 19.75& 19.87& 19.41& 19.58& 19.46& 19.47& 18.72& 19.29& 99.00& 21.78& 20.48& 19.83& 19.44& 19.17\\
   91& 351.26245117& 16.52570343&  0.024& 1& 99.00& 99.00& 21.90& 21.17& 20.25& 20.81& 20.70& 19.87& 19.49& 20.36& 19.56& 19.69& 19.63& 99.00& 19.79& 21.82& 20.60& 19.96& 19.81& 19.55\\
   92& 351.27828979& 16.76721191&  0.044& 1& 19.06& 18.34& 17.96& 17.76& 17.63& 17.57& 17.44& 17.36& 17.19& 17.24& 17.24& 17.11& 17.04& 16.98& 17.24& 19.02& 17.72& 17.33& 17.13& 17.03\\
   93& 351.28054810& 16.81988525&  0.029& 1& 19.76& 21.59& 19.68& 19.32& 19.33& 19.17& 18.97& 19.06& 18.78& 19.34& 18.74& 18.81& 18.52& 19.08& 19.11& 20.63& 19.61& 19.02& 18.82& 18.59\\
   94& 351.30871582& 16.44901848&  0.024& 2& 18.91& 19.99& 19.60& 19.39& 19.18& 19.07& 18.98& 18.66& 18.48& 18.47& 18.37& 18.25& 18.15& 17.96& 18.00& 19.73& 19.32& 18.69& 18.27& 18.02\\
   95& 351.31192017& 16.55031013&  0.017& 1& 19.12& 20.74& 19.24& 18.62& 18.41& 18.37& 17.88& 17.83& 17.64& 17.57& 17.59& 17.47& 17.42& 17.29& 17.33& 20.57& 18.61& 17.81& 17.50& 17.33\\
   96& 351.36566162& 17.25617027&  0.048& 7& 99.00& 20.54& 21.00& 20.06& 19.55& 19.86& 19.31& 19.08& 19.26& 19.51& 19.51& 19.36& 18.81& 19.88& 19.17& 20.47& 19.86& 19.60& 19.51& 19.08\\
   97& 351.40484619& 17.24501801&  0.037& 2& 20.47& 20.04& 19.57& 19.46& 19.07& 19.01& 18.73& 18.61& 18.37& 18.26& 18.23& 17.94& 17.97& 18.00& 18.11& 21.97& 19.22& 18.56& 18.25& 18.03\\
   98& 351.41052246& 16.57383156&  0.048& 3& 20.84& 20.53& 20.54& 19.65& 19.35& 19.23& 99.00& 18.92& 18.79& 18.55& 18.84& 99.00& 18.38& 18.14& 18.30& 20.70& 19.60& 18.88& 18.54& 18.24\\
   99& 351.41250610& 17.13726997&  0.045& 3& 20.05& 19.68& 19.11& 19.24& 18.84& 18.54& 18.43& 18.31& 18.14& 18.18& 17.83& 17.90& 17.80& 17.93& 17.88& 26.46& 19.04& 18.33& 17.97& 17.69\\
  100& 351.42208862& 16.75362396&  0.029& 1& 19.70& 20.91& 20.45& 19.74& 19.97& 19.86& 19.42& 19.17& 19.08& 19.25& 18.85& 99.00& 18.47& 18.67& 19.07& 21.26& 20.19& 19.20& 18.86& 18.47\\
  101& 351.42596436& 16.77386093&  0.039& 4& 19.45& 19.22& 20.35& 19.36& 19.23& 19.17& 18.87& 19.01& 18.88& 18.52& 18.62& 99.00& 18.30& 18.49& 18.30& 20.22& 19.36& 18.88& 18.56& 18.54\\
  102& 351.42956543& 16.78918457&  0.046& 3& 19.33& 23.81& 20.52& 20.13& 19.69& 19.60& 19.31& 19.47& 19.41& 19.18& 19.42& 99.00& 18.68& 18.80& 19.43& 21.38& 20.04& 19.38& 18.97& 18.86\\
  103& 351.43093872& 16.47374535&  0.048& 3& 19.44& 23.05& 20.24& 19.95& 19.51& 19.50& 19.02& 19.08& 18.80& 18.71& 18.94& 99.00& 18.46& 18.86& 18.50& 21.30& 19.84& 19.16& 18.86& 18.53\\
  104& 351.45364380& 16.79960823&  0.018& 2& 99.00& 99.00& 19.24& 19.30& 19.18& 19.10& 18.81& 18.77& 18.66& 18.29& 18.40& 99.00& 18.18& 17.97& 17.96& 20.85& 19.39& 18.67& 18.35& 18.08\\
  105& 351.45440674& 16.41203117&  0.029& 7& 19.82& 19.38& 19.05& 19.26& 19.08& 18.99& 18.90& 19.03& 18.79& 18.89& 18.90& 99.00& 18.67& 18.62& 18.37& 21.07& 19.22& 18.97& 18.79& 18.68\\
  106& 351.46267700& 16.88102722&  0.058& 3& 99.00& 20.35& 20.28& 19.93& 19.58& 19.59& 19.24& 19.39& 19.26& 19.57& 19.22& 99.00& 18.70& 19.15& 19.89& 23.66& 19.82& 19.35& 19.09& 19.06\\
\noalign{\smallskip}\hline
\end{tabular}
\end{table}

\section{Physical Properties of Regular Cluster A2589}

\subsection{Spatial Distribution}
The X-ray images obtained in the ROSAT \citep{buote96} and $Chandra$
\citep{buote04} observations support a picture that A2589 has a
smooth X-ray morphology and no sign of ongoing merger. The optical
map of cluster galaxies also shows no evidence of subclusters in
A2589 \citep{beers91}. The projected distribution of the 68 member
galaxies in sample I is given in Figure~10(a). In order to show the
overall morphology of galaxy two-dimension distribution, we superpose
the contour map of the surface density that has been smoothed by a
Gaussian window with $\sigma$ =3${\arcmin}$. It is easy to find that
A2589 concentration is rather compact: more than 50\% of bright known
member galaxies are located within a small circular region with
$r=0.5$ Mpc. The contour profiles tend to be more asymmetric at
larger radius, but no detached clumps (i.e. substructures) of
galaxies are detected. Figure~10(a) shows an elongation along the
north-south direction, which seems to agree with the X-ray brightness
profile and the orientation of central cD galaxy.

For the 174 member galaxies in sample II, the projected
distribution is also given in Figure~10(b). We superpose the
contour map of the spatial distribution that has been smoothed by
a Gaussian window with $\sigma =4\arcmin$. The filled circles denote the
68 member galaxies with known spectroscopic redshifts, and the
open circles denote the 103 newly-selected member galaxies. Though
the sample size increases by nearly 150\%, the profile of surface
density does not change a lot, with the same orientation and shape
in central region, and no discrete substructure is detectable at a
low surface density level of 0.14 arcmin$^{-2}$.

\begin{figure}[!h]
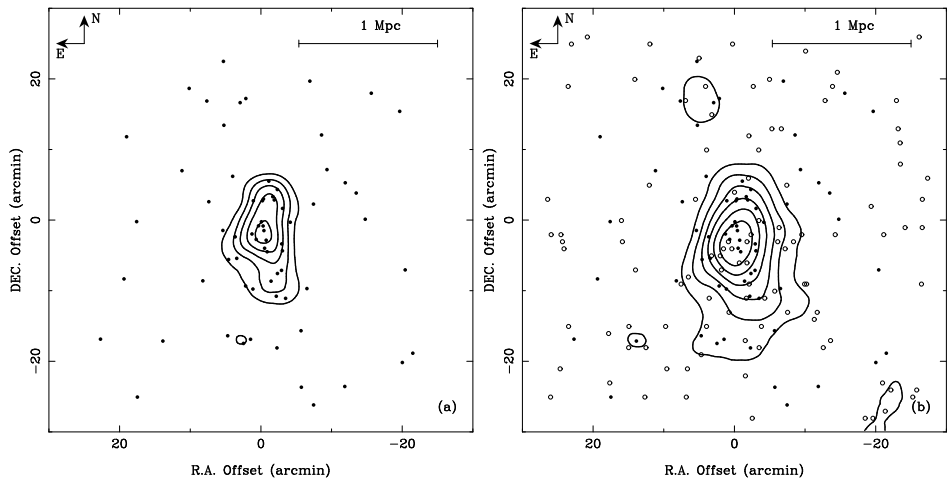
\centering
\includegraphics[width=62mm]{fig10a.eps}
\includegraphics[width=62mm]{fig10b.eps}
\caption{ \baselineskip 3.6mm Left: Spatial distribution for 68
spectroscopic member galaxies of A2589 in sample I. The contour
map of the surface density using the smoothing Gaussian window
$\sigma=3\arcmin$ is used. Right: Spatial distribution for 171 member
galaxies in sample II. The smoothing Gaussian window is
$\sigma$=4\arcmin. For both panels, the contour levels are 0.09, 0.14,
0.19, 0.24, 0.29, and 0.34 arcmin$^{-2}$, respectively.
 \label{fig10}}
\end{figure}

\subsection{Localized Velocity Structure}

Projected distribution of galaxies appears smooth and relaxed,
which might be due to projection effect. A true substructure
should be verified in the line-of-sight velocity space. The
$\kappa$-test has been commonly used for this purpose, which was
developed by \citet{colless96} for quantifying localized variation
in velocity distribution. A test statistic $\kappa_n$ is defined
to characterize the local deviation on the scale of the groups of
{\it n} nearest neighbors based on the Kolmogorov-Smirnov test
(KS-test). Larger $\kappa_n$ means greater possibility that the
local velocity distribution differs from the whole distribution.
The probability that $\kappa_n$ is larger than the observed value,
P($\kappa_n>\kappa_n^{obs}$), can be estimated by Monte Carlo
simulations by randomly shuffling velocities. Table 4 gives the
results of the $\kappa$-test for samples I and II, and the 10$^{3}$
simulations are made for all cases. A limit of
P($\kappa_n>\kappa_n^{obs}$) for substructure detection is 5\%,
corresponding to the $2\sigma$ significance. For the 68 member
galaxies in sample I, the probability P($\kappa_n>\kappa_n^{obs}$)
is found to be about 80\% (much more than the limit 5\%) in a wide
range of neighbor size, which strongly supports non-detection of
substructure. Even for the enlarged sample of 174 member galaxies,
the probability P($\kappa_n>\kappa_n^{obs}$) is still over 16\%
which is greater by two times than the limit 5\%. No substructures
are detected at more than $2\sigma$ significance on the basis of
sample II.

Figure~11 shows the bubble plots of the localized velocity
variation, using the neighbor size $n=9$ for both samples. The
bubble size for each galaxies is proportional
to $-$log[P$_{\rm KS}$(D$>$D$_{\rm obs}$)]. Therefore, the larger
bubbles indicate a greater difference between local and overall
velocity distributions. There is no prominent bubble clustering in
the core region of A2589. For the 68 spectroscopic galaxies in
sample I, the central bubbles are very tiny, indicating that the
local velocity distribution is in well accordance with the overall
one. Even for the enlarged sample II, bubble clustering at center
is still neglectable. A close comparison between the projected
distribution (in Figure~10) and bubble plot (in Figure~11) shows
that A2589 is a well-relaxed cluster with no dynamic substructures
detected both in 2-d mapping and in radial velocity space.

An elongation along the north-south direction has been unveiled in
Figure~10. To make certain of whether there exists a substructure
located about 10 arcmin south to the main concentration, we further
apply a currently favored technique of mixture modeling, namely the
KMM algorithm, to the samples I and II. The KMM is a
maximum-likelihood algorithm which assigns objects into groups and
assesses the improvement in fitting a multi-group over a single group
model (Ashman et al. 1994). Based on the three-dimension data (i.e.,
projected positions and radial velocities of member galaxies), we set
the initial positions: (0.0, 0.0) for the main concentration and
(-2.0, -11.0) for the possible substructure, and the initial mean
velocities are 12050 \kms for two clumps. The KMM algorithm gives
following optimum two-group solutions: only three galaxies in sample
I are assigned as the members of southern substructure, and only five
galaxies in sample II are allocated to the substructure. This means
the southern anormal feature in projected mapping is not a
significant substructure, which supports the conclusion drawn by the
$\kappa$-test that no significant dynamic substructures are found in
A2589.

\begin{figure}[t]
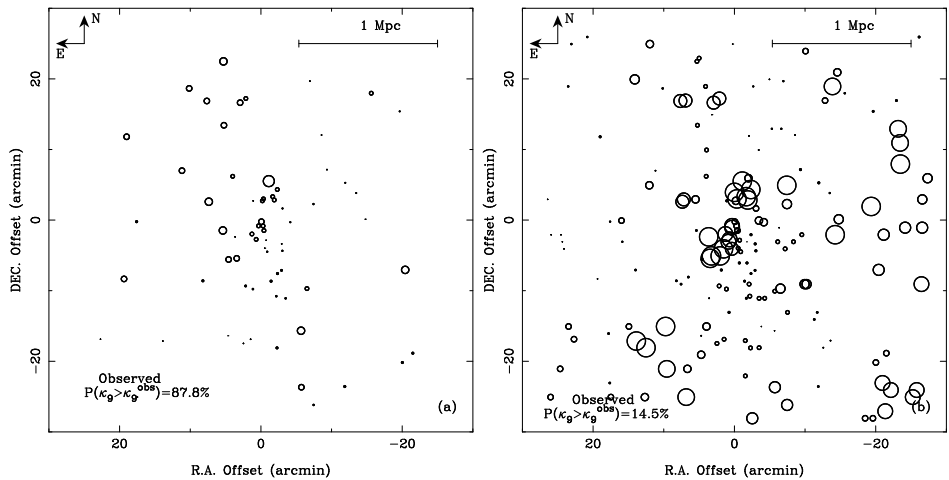

\centering
\includegraphics[width=62.0mm]{fig11a.eps}
\includegraphics[width=62.0mm]{fig11b.eps}
\caption{ \baselineskip 3.6mm (a) Bubble plot showing the localized
variation for groups of the 9 nearest neighbors for (a) 68
galaxies in sample I; (b) 174 galaxies in sample II  \label{fig11}}
\end{figure}

\begin{table}[h!!]
\vspace{0pt} \centering
\begin{minipage}{115mm}
\caption[]{Results of $\kappa$-test for member galaxies in samples
I and II}
\end{minipage}

\begin{tabular}{ccccccc}
\hline\noalign{\smallskip}
{Neighbor size} & Sample I & Sample II \\
$n$ & $P$($\kappa_n>\kappa_n^{obs}$) & $P$($\kappa_n>\kappa_n^{obs}$)\\
\noalign{\smallskip}   \hline \noalign{\smallskip}
 6.........& 84.8\% & 17.8\% \\
 7.........& 85.9\% & 21.4\% \\
 8.........& 81.9\% & 13.8\% \\
 9.........& 87.8\% & 14.5\% \\
10.......  & 85.2\% & 21.7\% \\
11.......  & 85.9\% & 31.2\% \\
12.......  & 79.9\% & 49.2\% \\
\noalign{\smallskip}   \hline\noalign{\smallskip}
\end{tabular}
\end{table}

\subsection{Luminosity Function}
The luminosity function (LF) is a fundamental tool to analyze the
properties of galaxies in a cluster. Usually, a Gaussian function can be
used to describe the LF of bright galaxies with $i<19^m.0$ \citep{binggeli88}
$ N(M)\sim exp[-(M-\mu)^{2}/2\sigma^{2}]$,
where $\mu$ is the characteristic absolute magnitude,
and $\sigma$ is the dispersion. For the faint galaxies with $i>19^m.0$,
a Schechter function is needed to give a better fit \citep{schechter76}:
$N(M) =
\phi^{\ast}[10^{-0.4(M-M^{\ast})}]^{(\alpha+1)}e^{-10^{-0.4(M-M^{\ast})}}$,
where $\phi^{\ast}$, $M^{\ast}$, and $\alpha$ are the
normalization parameter, the characteristic absolute magnitude, and the
slope at faint end, respectively. For the cluster galaxies, a single
Schechter function can not give a perfect description for the LFs
\citep{driver94, mohr96, trentham02, yang04}.
\citet{ferguson91} performed a LF fitting with these two functions.

For comparing with previous results, the BATC magnitudes are
converted into the conventional Kron-Cousins magnitudes via the
equations in \citet{zhou03}. The $M_{R}$ magnitudes in sample II
cover a range from $-22.0$ to $-16.5$ mag, and a turn-off point at
$M_{R}=-18.75$ is found in magnitude distribution. We use the
combination of above two functions to describe the LFs of galaxies in
A2589: a Gaussian function for the bright galaxies, and a Schechter
function for the faint galaxies. The fitting results in a Gaussian
function with $\mu=-20.0^{+0.02}_{-0.18}$,
$\sigma$=$-1.34^{+0.02}_{-0.02}$, and a Schechter function with
$M^{\ast}=-17.8^{+0.35}_{-0.01}$, $\alpha=-0.53^{+0.20}_{-0.04}$. The
combination of two functions gives a good description for the LFs of
A2589 (see Figure~12).

Though the uncertainty in member selection might be larger for faint
galaxies, the LF peak at $M_{R} \sim -20$ and dip at $M_{R} \sim -19$
might be real because the BATC photometry is capable for detecting
the galaxies brighter than 19.5 mag. With the clear peak at $M_{R}
\sim -20$, the bright part LF of A2589 looks similar with the LF of
rich clusters, which have a bump in bright part \citep{biviano95}.
Our turn-off magnitude $M_{R}=-18.75$ is similar to the LFs in A963
\citep{driver94}, Coma \citep{thompson93}, and A2554 \citep{smith97}.
For the LF slope at faint parameter $\alpha$, A2589 has a flat slope
than many rich clusters, which might be due to the incompleteness in
selection of faint members.

Following the definition of dwarf-to-giant ratio in \citet{driver98},
we define the faint-to-bright ratio (FBR) as the count ratio of faint
galaxies to bright galaxies, $FBR$=${\sum N(M_R>-18.75)}$ $/{\sum
N(M_{R}<-18.75)}$. The overall $FBR$ is about 0.84. It is found that
the $FBR$ varies with clustercentric radius: $FBR=0.44$ in central
region ($R\leq$0.5Mpc), and $FBR = 1.23$ in outer region ($R
>$ 0.5Mpc). The increasing tendency of the FBR along radial distance
is presented in Figure~13. Obviously, the LF faint end is dominated
by faint/dwarf galaxies, which can be interpreted by the `dwarf
population density' relation \citep{phillipps98}: dwarves are more
common in  the low density environment. This segregation may
originate from initial conditions of formation of dwarf galaxies,
where low luminosity galaxies are only now in-falling into clusters
\citep{croton05}, or the galaxies may have suffered the processes
internal to clusters, such as tidal disruption and galaxy harassment
\citep{mastropietro05, aguerri04, aguerri05}, consequently dimmed.

\begin{figure}[t] \centering
\includegraphics[width=80mm]{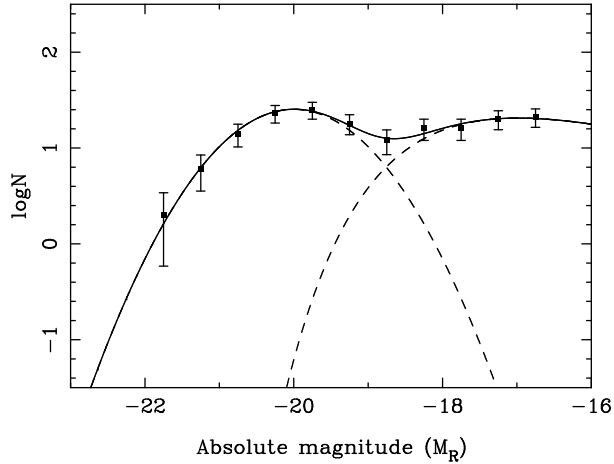}
\caption{Observed luminosity function of R band. The dashed lines
indicate the components of the Gauss and Schechter function, the
solid lines are their sum. \label{fig12}}
\end{figure}

\begin{figure}[h] \centering
  \includegraphics[width=80mm]{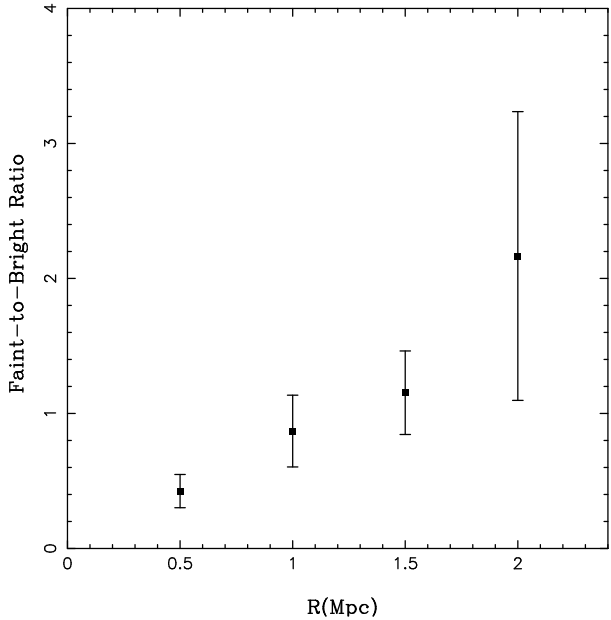}
\caption{The faint-to-bright ratio ($FBR$) as a function of
clustercentric distance of A2589. The mean FBRs are computed in
annuli of 0.5Mpc width . \label{fig13}}
\end{figure}

\subsection{Star Formation Properties of Cluster Galaxies}
The star formation histories of the member galaxies can provide
important clues for understanding the evolution of their host
cluster. For a cluster galaxy, both the cluster-scale gravity
environment and the galaxy-scale interaction may have influenced
the physical processes concerning star formation. Therefore, it is
interesting to observe the systematic tendency of the star
formation properties for the galaxies in a cluster.

With an evolution synthesis model, PEGASE (version 2.0,
\citet{fioc97,fioc99}), the star formation properties of A2589 are
investigated. We assume a \citet{salpeter55} initial mass function
(IMF) and a star formation rate (SFR) in exponentially decreasing
form, $SFR(t) \propto e^{-t/{\tau}}$, where the time scale $\tau$
ranges from 0.5 to 30.0 Gyr. To avoid the degeneracy between age and
metallicity in the model, we use the same age of 12.86 Gyr for all
the member galaxies in A2589, responding to the age of the first
generation stars at $z_c$ =0.0414. A zero initial metallicity of
interstellar medium (ISM) is assumed. Firstly, a series of rest-frame
modeled spectra at {\it z}=0 with various star formation histories
are generated by running the PEGASE code. Then we shift them to the
observer's frame for a given redshift, and then convolve with the
transmission functions of the BATC and SDSS filters. As a result, we
obtain the template SED library containing the BATC and SDSS
photometric bands.

Based on the template SED library, we search for the minimal
$\chi^{2}$ fit of the observed SEDs of 68 bright member galaxies
with known spectroscopic redshifts.
The SFR time scale ($\tau$), mean ISM metallicity (Z$_{\rm ISM}$), and
the mean stellar age ($t_{\star}$) weighted by mass and light
can be achieved.


Figure~14 shows the star formation properties as a function of the
projected radius ({\it R}) for 68 member galaxies in sample I.
The dash lines denote the best linear fitting
results. Panel (a) shows that the SFR time scale is shorter in the
inner region than that in the outer region. This result
agrees with the morphology-density relation pointed out by
\citet{dressler80}, which can be well explained in the context of
hierarchical cosmological scenario \citep{poggianti04}. Panel (b)
presents that the outlier galaxies have higher probability to
have a lower metallicity, which is consistent
with the pictures that more massive galaxies form fractionally more
stars in a Hubble time than the low-mass counterparts, and metals
are selectively lost from the faint galaxies with shallow
potential wells via galactic winds \citep{tremonti04}. Panels (c)
and (d) show that the galaxies in the core region tend to possess
older stellar population with longer mean stellar ages weighted by
either mass or light. Comparison between panels (c) and (d) shows
that the gradient of the light-weighted stellar age is steeper.
After the evolution of galaxies, stellar mass are dominated by
old stellar population with dimmed light. On the other hand, current
star formation rate in a cluster is mainly contributed by the late-type galaxies in the outer region, and younger stellar population has a greater weight
in mean age calculation. It is reasonable that the stellar ages
weighted by bolometric luminosity tend to be younger for the outlier galaxies.

\begin{figure}[!t]
\centering
\includegraphics[width=120mm]{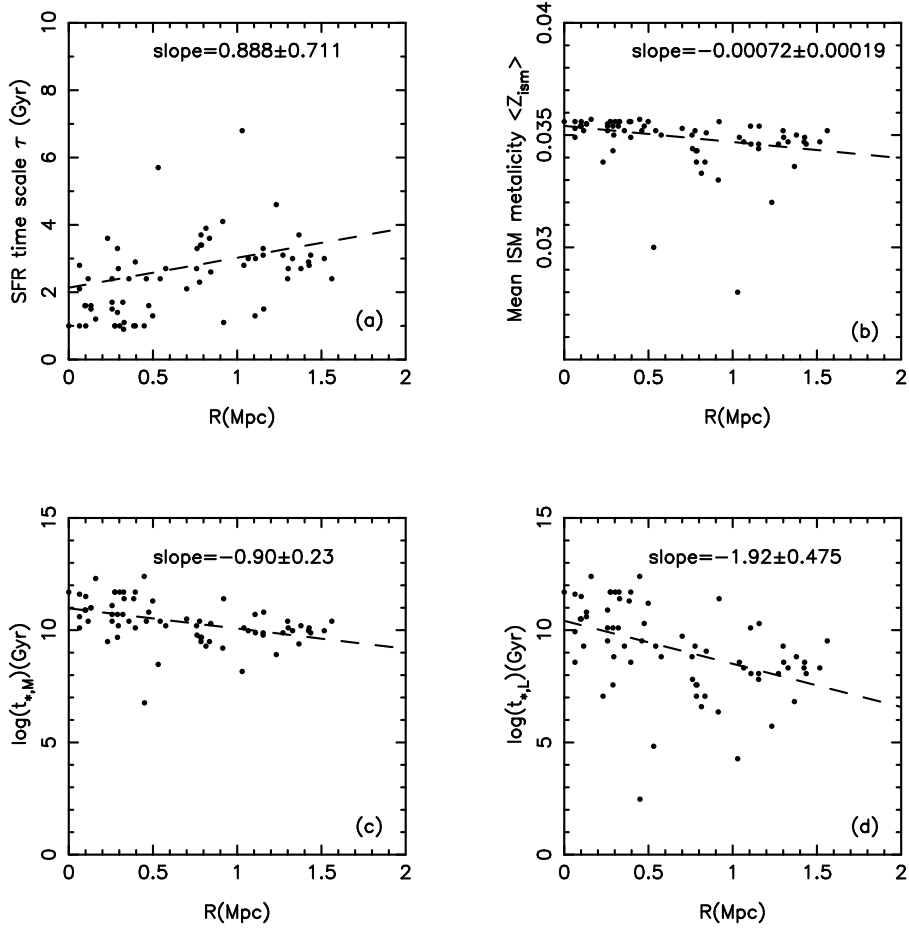}
\caption{Star formation properties for the galaxies with known
{\it z$_{sp}$} in A 2589 as a function of the project radius R.
The star formation properties are the SFR time scale $\tau$, mean
ISM metallicity, and the mean stellar ages weighted by mass and
light.\label{fig14}}
\end{figure}

Figure~15 gives the SFR time scale $\tau$ and the ISM metallicities
as a function of absolute magnitude ({\it M$_{R}$}). The high
luminosity galaxies tend to have shorter time scales. For the bright
and massive cluster galaxies in the core region of A2589, their
star formation activities might have been reduced by many different
physical process via environmental effects, such as galaxy-galaxy
interaction, harassment, gas stripping, strangulation
\citep{poggianti04,yuan05}, which results in a short time scale of
star formation.

\begin{figure} [ht]
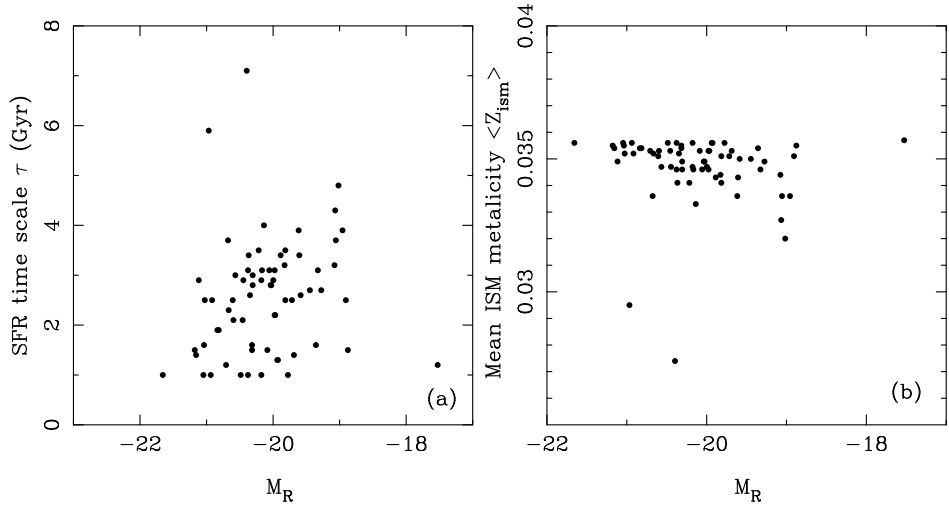

\centering
\includegraphics[width=62mm]{fig15a.eps}
\includegraphics[width=62mm]{fig15b.eps}
\caption{The SFR time scale $\tau$ and mean ISM metallicities for
the galaxies with known {\it z$_{sp}$} in A2589 as the function of
magnitude in R band.\label{fig15}}
\end{figure}

\section{Summary}

We present our multicolor optical photometry for the nearby regular cluster
of galaxies A2589, based on the BATC 15-intermediate bands system
and SDSS photometric data. The SEDs in 15 bands are obtained
for more than 5000 sources detected from $\sim$1 deg$^{2}$ of the BATC images.
After cross-identifying the BATC sources with the SDSS photometric catalog,
1199 galaxies are extracted. An interpolation method is performed to make
zero-point correction, the combined 20-band SEDs for 1199 galaxies are archived.
Then, a photometric redshift technique and the CM relation of early-type galaxies
are applied to select fainter member galaxies. As a result, 106 galaxies
are selected as the faint member galaxies. Combining 68 member galaxies
with known spectroscopic redshifts (i.e., sample I) , we obtain an enlarged
sample of 174 member galaxies (i.e., sample II).

The projected distribution shows no prominent clumps. The contour of
surface density shows an north-south elongation, which agrees with
the X-ray brightness profile and the orientation of central cD
galaxy, NGC~7647. Subsequential $\kappa$-test also indicates no
substructure in the galaxy cluster A2589, which agrees with the X-ray
images by $ROSTAT$ \citep{buote96} and {\it Chandra} \citep{buote04}.
Our conclusion is that A2589 is a well-virialized and relaxed system.

The luminosity function of member galaxies in A2589 shows a peak at $M_R \sim -20.0$
and a dip at M$_{R}$ $\sim -19.0$. Compared with other clusters, the turn-off point
seems to be independent on the richness and dynamic stage. The faint-bright-ratio (FBR)
increases monotonously along clustercentric distance, indicating that
the faint and dwarf galaxies tend to be located in the outer region of cluster.

The star formation properties of cluster galaxies unveils an
environmental influence on evolution of A2589. Bright and massive
galaxies in the core region are found to have shorter SFR time
scales,  longer mean stellar ages, and higher mean ISM metallicities,
while the outlier galaxies are likely to have smaller stellar ages,
and longer SFR time scales.  These results can be well interpreted by
the existing correlations, such as the morphology-density relation,
the luminosity-metallicity relation, and the mass-metallicity
relation.

\acknowledgments We thank the anonymous referee for his/her
invaluable comments and suggestions. This work was funded by the
National Natural Science Foundation of China (NSFC) (Grant
Nos.~10778618, 10633020, 10873016, 10603006, and 10803007), and by
the National Basic Research Program of China (973 Program) (Grant
No.~2007CB815403). This research has made use of the NED, which is
operated by the Jet Propulsion Laboratory, California Institute of
Technology, under contract with the National Aeronautics and Space
Administration. We would like to thank Prof. Kong, X. and Cheng,
F.-Z. at the University of Science and Technology of China for the
valuable discussion.



\clearpage

\end{document}